\documentclass[aps,preprint,nofootinbib,floatfix]{revtex4}
\usepackage{graphicx,color}
\usepackage[latin1]{inputenc}
\usepackage{amsmath,amssymb}
\usepackage{hyperref}
\usepackage{epstopdf}
\usepackage{slashed}
\usepackage{soul}
\usepackage{amssymb}
\usepackage[normalem]{ulem}
\usepackage[caption=false]{subfig}
\usepackage{booktabs}
\usepackage{diagbox}
\usepackage{makecell}
\usepackage{multirow}
\newcommand{\lsim}{\mathrel{\mathop{\kern 0pt \rlap
  {\raise.2ex\hbox{$<$}}}
  \lower.9ex\hbox{\kern-.190em $\sim$}}}
\newcommand{\gsim}{\mathrel{\mathop{\kern 0pt \rlap
  {\raise.2ex\hbox{$>$}}}
  \lower.9ex\hbox{\kern-.190em $\sim$}}}

\newcommand{\sv}{\ensuremath{\langle\sigma v\rangle}}

\newcommand{\sigsip}{\ensuremath{\sigma^{\rm{SI}}_p}}

\newcommand{\mev}{\ensuremath{\,\mathrm{MeV}}}
\newcommand{\gev}{\ensuremath{\,\mathrm{GeV}}}
\newcommand{\tev}{\ensuremath{\,\mathrm{TeV}}}

\def  \bcen   {\begin{center}}
\def  \ecen   {\end{center}}
\def  \beq    {\begin{equation}}
\def  \eeq    {\end{equation}}
\def  \beqa   {\begin{eqnarray}}
\def  \eeqa   {\end{eqnarray}}

\def\bea{\begin{eqnarray}}
\def\eea{\end{eqnarray}}

\begin{document}

\title{Light Thermal Dark Matter Beyond $p$-Wave Annihilation in Minimal Higgs Portal Model}
\author{Yu-Tong Chen$^{a,b}$}
\author{Shigeki Matsumoto$^{c}$}
\author{\\Tian-Peng Tang$^{a}$}
\author{Yue-Lin Sming Tsai$^{a,d}$}
\author{Lei Wu$^{b}$}

\affiliation{$^a$Key Laboratory of Dark Matter and Space Astronomy, 
   Purple Mountain Observatory, Chinese Academy of Sciences, Nanjing 210033, China}
\affiliation{$^b$Department of Physics and Institute of Theoretical Physics, 
Nanjing Normal University, Nanjing, 210023, China }
\affiliation{$^c$Kavli IPMU (WPI), UTIAS, U. Tokyo, Kashiwa 277-8583, Chiba, Japan}
\affiliation{$^d$School of Astronomy and Space Science, University of Science and Technology of China, Hefei, Anhui 230026, China}

\date{\today}

\begin{abstract}
This study explores a minimal renormalizable dark matter (DM) model, incorporating a sub-GeV Majorana DM and a singlet scalar particle $\phi$. Using scalar and pseudo-scalar interactions (couplings $c_s$ and $c_p$), we investigate implications for DM detection, considering $s$-wave, $p$-wave, and combined ($s$+$p$ wave) contributions in DM annihilation cross-section, as well as loop-correction contributions to DM-nucleon elastic scattering. Identifying a broad parameter space ($10\mev < m_\chi \lesssim m_\phi$) within the $2\sigma$ allowed region, we explore scenarios ($\left|c_s\right|\gg \left|c_p\right|$, $\left|c_s\right|\ll \left|c_p\right|$, and $\left|c_s\right|\approx \left|c_p\right|$). 
We find that (i) a non-zero pseudo-scalar coupling alleviates direct detection constraints as a comparison with the previous pure scalar coupling case; (ii) CMB observations set stringent limits on pseudo-scalar interaction dominant cases, making $s$-wave annihilation viable only for $m_\chi>1\gev$; (iii) the preferred $\phi$-resonance region can be tested in the future indirect detection experiments, such as e-ASTROGAM.

\end{abstract}

\maketitle

\section{Introduction \label{sec:intro}}

The nature of dark matter (DM) is one of the most fundamental problems in modern physics, 
and revealing its physical properties could lead to substantial progress in fundamental physics.
Among the various candidates for DM, the thermal DM model stands out for its natural explanation of the observed DM relic abundance. 
This model relies on the freeze-out mechanism, a powerful framework that not only sheds light on the DM nature, 
but also provides a successful explanation for the history of Big Bang Nucleosynthesis (BBN).

The mass range of thermal DM is theoretically predicted to span from a few MeV to several hundreds of TeV~\cite{Boehm:2003bt,Boehm:2002yz,Murayama:2009nj,Hambye:2009fg,Boehm:2013jpa}.  
The upper and lower bounds of DM mass are both related to the constraints of the relic density and are based on different assumptions. 
The mass upper bound comes from the joint constraint of the unitarity condition for $s$-wave annihilation~\cite{Griest:1989wd}, and the relic density. 
On the other hand, the lower bound of the DM mass depends on the particles involved in DM annihilation.
If DM only annihilates into a pair of standard model (SM) particles mediated by any SM particle, then the relic density constraint requires DM mass to be heavier than $2\gev$, called Lee-Weinberg bound~\cite{Lee:1977ua}. 
However, if the DM annihilates to a pair of new light scalar particles, as discussed in this work, 
the scalar particles must be larger than MeV to be consistent with the allowed relativistic degrees of freedom in the BBN era~\cite{Depta:2019lbe,Cyburt:2002uv,Jedamzik:2009uy,Hufnagel:2018bjp,Kawasaki:2020qxm,Depta:2020zbh}.

Thus, the DM mass has to be heavier than $\mathcal{O}(10\mev)$ to simultaneously fulfill the relic density constraint and the allowed relativistic degrees of freedom.  

Direct detection, a primary strategy for weakly interacting massive particle (WIMP) DM exploration through scattering interactions with nuclei, is constrained to DM masses above a few GeV due to the required kinetic energy for nuclear recoil through elastic scattering. Recent experiments like XENON~\cite{XENON:2023cxc}, LZ~\cite{LZ:2022lsv}, and PandaX-4T~\cite{PandaX-4T:2021bab} have ruled out a large portion of the parameter space for conventional GeV-scale WIMP DM, leaving a vast region of sub-GeV thermal DM untested. Detecting nuclear recoil energy in sub-GeV thermal DM using conventional methods poses challenges. Methods for DM direct detection with masses ranging from eV to $100\mev$ are to search for electron recoils, with active pursuit by various groups~\cite{SuperCDMS:2018mne, DAMIC:2019dcn, PandaX-II:2021nsg, CDEX:2022kcd, SENSEI:2023zdf}. This study aims to explore the unknown parameter space of a minimal renormalizable DM model including a sub-GeV Majorana DM and a singlet scalar particle~\cite{Matsumoto:2018acr}. In this Higgs portal DM model, the DM-electron interaction is suppressed by the small electron Yukawa coupling and a tiny mixing angle, which simply escapes from the experimental constraints involving electron recoils.

In the early universe, charged particles interacted with CMB photons, altering their black-body spectrum~\cite{Chluba:2011hw}.
After recombination, injected energy affected gas temperature and ionization fraction, leaving imprints on CMB temperature and polarization power spectra~\cite{Adams:1998nr,Chen:2003gz,Padmanabhan:2005es,Slatyer:2009yq}.
CMB measurements, evolving from pre-WMAP to the Planck satellite era, independently constrain DM annihilation cross-sections, particularly in the sub-GeV mass range.
Other indirect detection experiments like Fermi-LAT and AMS-02 lose sensitivity in the low-mass regime due to instrumental thresholds~\cite{Ackermann:2015zua,Hoof:2018hyn,Oakes:2019ywx,Boudaud:2016mos,Roszkowski:2017nbc}.
Overall, CMB serves as a distinct and complementary probe for understanding thermal DM, crucial for low-mass DM and models with non-velocity-suppressed $s$-wave annihilation~\cite{Slatyer:2009yq,Slatyer:2015jla,Steigman:2015hda,Planck:2018nkj,Cang:2020exa,Kawasaki:2021etm,Matsumoto:2022ojl}.
Beyond $s$-wave annihilation, several mechanisms can simultaneously meet CMB constraints and yield the correct relic abundance, such as $p$-wave DM annihilation~\cite{Diamanti:2013bia,Bondarenko:2019vrb,Croon:2020ntf,Ding:2021sbj,Binder:2022pmf,Belanger:2024bro,Siegert:2024hmr}, forbidden DM mechanisms~\cite{DAgnolo:2015ujb,DAgnolo:2020mpt,Hara:2021lrj,Wojcik:2021xki,Cheng:2022esn}, self-interaction DM~\cite{Boehm:2003hm,Lin:2011gj,Chu:2016pew,Tulin:2017ara}, and other alternatives~\cite{Mizuta:1992qp,Pospelov:2007mp,Feng:2008mu,Feng:2010zp,Hochberg:2014dra,Hochberg:2014kqa,Bernreuther:2020koj,Guo:2023kqt}.

In this study, our goal is to identify a parameter space of the minimal Higgs portal DM model that mitigates constraints from direct detection and CMB, while remaining testable in future indirect detection experiments like e-ASTROGAM~\cite{e-ASTROGAM:2016bph,DeAngelis:2021esn}. 
The DM candidate in this minimal model is singlet under the SM gauge group and interacts with the SM sector via a singlet scalar~\cite{Matsumoto:2018acr}.
We conduct a comprehensive Markov chain Monte Carlo (MCMC) analysis, considering constraints from cosmology, astrophysics, direct detection experiments, collider experiments, DM self-interaction, and CMB observations. 
Our findings indicate that thermal light DM, annihilating through an $s$-wave process, fails to simultaneously satisfy relic density and CMB constraints, 
if DM mass is lighter than $1\gev$. 
Although $p$-wave annihilation may evade CMB constraints, the small annihilation cross-section poses a challenge for future indirect detection experiments. Fortunately, upcoming experiments, such as e-ASTROGAM, COSI~\cite{Tomsick:2019wvo,Caputo:2022dkz}, and GECOO~\cite{Orlando:2021get,Coogan:2021rez}, designed for the MeV-Gap, offer promising prospects to explore $\phi$-resonance annihilation regions of sub-GeV DM and enhance sensitivity by 2-3 orders of magnitude compared to current limits~\cite{Essig:2013goa,Boddy:2015efa}.

The remaining sections of the paper are organized as follows. 
In Sec.~\ref{sec:Model}, we present the minimal Higgs portal DM model featuring a light Majorana DM coupled with a singlet scalar. 
We provide a comprehensive overview of the interactions and the decay of this new scalar. 
In Sec.~\ref{sec:Constraints}, we outline all the constraints incorporated into our likelihood, along with a detailed discussion of the updated constraints. 
The outcomes of our likelihood analysis are presented in Sec.~\ref{sec:result}. 
Finally, we summarize our findings in Sec.~\ref{sec:summary}.

\section{a minimal Higgs portal DM model}
\label{sec:Model}

We consider an SM singlet Majorana DM $\chi$ interacting with the SM sector via an SM singlet real scalar boson $\Phi$~\cite{Matsumoto:2018acr}. For DM mass $m_\chi$, the minimal but renormalizable Lagrangian can be written as 
\begin{equation}
	\mathcal{L} =
	\mathcal{L}_{\rm SM} + \frac{1}{2} \bar{\chi} (i\slashed{\partial} - m_{\chi}) \chi + \frac{1}{2} (\partial \Phi)^2
	- \frac{c_s}{2} \Phi \bar{\chi} \chi - \frac{c_p}{2} \Phi \bar{\chi} i \gamma_5 \chi
	-V(\Phi,H), 
	\label{eq: lagrangian}
\end{equation}
where $\mathcal{L}_{\rm SM}$ and $H$ are the SM Lagrangian and SM Higgs doublet, respectively. 
The coupling coefficients of DM with scalar and pseudo-scalar are denoted as $c_s$ and $c_p$.
To preserve a stable DM, $\chi$ is the $Z_2$-odd particle, while other particles are $Z_2$-even.  The scalar potential of the model is composed of $V(\Phi, H) \equiv V_\Phi(\Phi) + V_{\Phi H}(\Phi, H)$ and $V_H(H)$, 
where $V_H(H)$ is the SM Higgs potential. 
Their explicit forms are 
\begin{eqnarray}
	V_H(H) &=& \mu^2_H H^{\dagger} H + \frac{\lambda_H}{2} (H^{\dagger} H)^2,
	\nonumber \\
	V_{\Phi}(\Phi) &=& \mu^3_1 \Phi + \frac{\mu^2_\Phi}{2} \Phi^2 + \frac{\mu_3}{3!} \Phi^3 + \frac{\lambda_\Phi}{4!} \Phi^4,
	\nonumber \\
	V_{\Phi H}(\Phi, H) &=& A_{\Phi H} \Phi H^{\dagger} H + \frac{\lambda_{\Phi H}}{2} \Phi^2 H^{\dagger} H,
	\label{eq: potential}
\end{eqnarray}
where $\lambda_i$s are dimensionless coupling constants of quartic terms while others $\mu_{1,2,3}$ and $A_{\Phi H}$ are dimensional couplings for quadratic and cubic scalar interactions.

\subsection{The properties of the scalar particles}
\label{sec:scalar}

In this subsection, we summarize the properties of the new scalar particle $\phi$ and the SM Higgs, 
including their masses, interactions, and decay channels.

\subsubsection{Masses of the scalars}
\label{sec:mass}

We denote the vacuum expectation value (VEV) of $H$ and $\Phi$ are $v_H$ and $v_\Phi$, respectively. 
By expanding $H=[0,(v_H+h')/\sqrt{2}]^T$ and $\Phi=v_\Phi+\phi'$, 
we then rewrite the mass matrix of the scalars with the flavor basis ($h'$, $\phi'$) as 
\begin{eqnarray}
    U
	\left( \begin{matrix} m^2_{h'h'} & m^2_{h'\phi'} \\ m^2_{h'\phi'} & m^2_{\phi'\phi'} \end{matrix} \right)
	U^\dagger
	=
	\left( \begin{matrix} m_h^2 & 0 \\ 0 & m_\phi^2 \end{matrix} \right),
	\qquad
	U
	=
	\left( \begin{matrix} \cos \theta & - \sin \theta \\ \sin \theta & \cos \theta \end{matrix} \right),
	\label{eq: mass matrix}
\end{eqnarray}
where we define the $m^2_{h'h'} = \lambda_H v_H^2$, $m^2_{h'\phi'} = A_{\Phi H} v_H$, and 
$m_{\phi'\phi'}^2 = \mu_\Phi^2 + \lambda_{\Phi H} v_H^2/2$. 
The mixing angle $\theta$ in the matrix $U$ controls the magnitude of interactions 
between the mediator particle $\phi$ and SM particles.
Note that the physical mass $m_h$ and $m_\phi$ can be obtained by the eigenvalue of the mass matrix. 
In addition, $\sin\theta$ is a common variable used in the Higgs portal DM model. 
Hence, instead of using $A_{\Phi H}$ and $\lambda_{\Phi H}$, we take $m_\phi$ and $\sin\theta$ as our model inputs.

\subsubsection{Scalar interactions and DM annihilations}

The interactions between the scalar sector ($h$ and $\phi$) with DM sector are
\begin{equation}
	\mathcal{L}_{\rm int} \supset
	- \frac{\cos\theta}{2}(c_s\,\phi\,\bar{\chi}\chi+c_p\,\phi\,\bar{\chi} i \gamma_5 \chi)
	+ \frac{\sin\theta}{2}(c_s\,h\,\bar{\chi}\chi+c_p\,h\,\bar{\chi} i \gamma_5 \chi).
\end{equation}
Clearly, the interactions between $h$ and $\chi$ are suppressed by $\sin \theta$. 
On the other hand, $\cos\theta$ is roughly equal to 1, and $\phi$ interacts with the dark sector only depending on $c_s$ and $c_p$.

We can rewrite the triple and quartic scalar interactions with the mixing angle, 
\begin{eqnarray}
\label{eq: h phi terms}
	\mathcal{L} &\supset&
	- \frac{c_{h h h}}{3!} h^3 - \frac{c_{\phi h h}}{2} \phi h^2
	- \frac{c_{\phi \phi h}}{2} \phi^2 h - \frac{c_{\phi \phi \phi}}{3!} \phi^3,
	\nonumber \\
    &&
    - \frac{c_{h h h h}}{4!} h^4 - \frac{c_{\phi h h h}}{3!} \phi h^3 - \frac{c_{\phi \phi h h}}{4} \phi^2 h^2
	- \frac{c_{\phi \phi \phi h}}{3!} \phi^3 h - \frac{c_{\phi \phi \phi \phi}}{4!} \phi^4,
\end{eqnarray}
where all the relevant couplings can be found in appendix~\ref{app: scalar interactions}. 
We would like to note that the couplings $c_{\phi \phi h}$ and $c_{\phi \phi h h}$ are not suppressed even when the mixing angle is very suppressed (namely, $|\sin \theta| \ll 1$), as they originate in the quartic term $\Phi^2 |H|^2$.

The new scalar boson $\phi$ plays an important role in DM annihilation. 
Compared with Ref.~\cite{Matsumoto:2018acr}, our work considers the non-zero pseudo-scalar coupling $c_{p}$. 
The newly introduced interaction may induce an $s$-wave DM annihilation cross-section, which remains unsuppressed in the present universe. 
This aspect has been explored in Refs.~\cite{Matsumoto:2014rxa, Berlin:2015wwa, Matsumoto:2016hbs, DiazSaez:2021pmg} for investigating various phenomenological possibilities~\cite{Pospelov:2007mp, LopezHonorez:2012kv, Heikinheimo:2013xua, Ghorbani:2014qpa, Kim:2016csm,Esch:2013rta, Esch:2014jpa, Ipek:2014gua, Kahlhoefer:2017umn, Baek:2017vzd, Arcadi:2017wqi, Bell:2017irk, Ghorbani:2018pjh, Banerjee:2021hfo}.
Therefore, the two main annihilation channels are: 
\begin{itemize}
    \item $\chi\chi\to {\rm SM} +{\rm SM}$\\
The $s$-channel with $\phi$-exchange is dominant in the relevant Feynman diagrams.  
If $m_\chi\approx m_\phi/2$, the resonance mechanism takes over the annihilations. 
In principle, resonance annihilation can happen at any time. 
We will show that the resonance mechanism can be detected in future MeV gamma-ray telescopes but 
fulfill the relic density and escape from the CMB constraints if DM resonance annihilation appears in the present Universe.

To understand the velocity dependence of the DM annihilation cross-section,    
we can expand the cross-section of DM annihilating to a pair of SM particles as 
$\left \langle \sigma v \right \rangle \propto c_{p}^{2}a +(c_{s}^{2}b_{1}+c_{p}^{2}b_{2} )v^{2}$, 
where the coefficients $a$, $b_{1}$ and $b_{2}$ are determined by the $m_{\chi},\; m_{\phi},\; \sin \theta$ and the mass of the SM particles. 
Clearly, if $c_{p}=0$ but $c_s \neq 0$, the cross-section $\left \langle \sigma v \right \rangle$ remains $v^2$-dependence, 
namely $p$-wave component. 
On the contrary, in the case of $c_{p}\neq 0$ and $c_s=0$, the expansion of $\left \langle \sigma v \right \rangle$ posses both 
$s$-wave and $p$-wave contributions, but the $s$-wave one can be enhanced by a small velocity.

It is important to clarify that the aforementioned expansion and velocity-dependence are not applicable to the $\phi$ resonance. 
As demonstrated in Ref.~\cite{Abdughani:2021oit} (see Eq.~(13) and Table I therein), the resonance peak and velocity-dependence strongly depend on the numerator of the cross-section. 
Given these considerations, we will address the $\phi$ resonance as a separate case in this study.

    \item $\chi\chi\to \phi +\phi$\\
    There are two diagrams of this channel. The first one is $t$-channel with $\chi$-exchange and the second one is $s$-channel with $\phi$ 
    exchange. The former usually relates to the forbidden DM scenarios ($m_\phi \gsim m_\chi$) or secluded DM scenarios ($m_\phi \ll m_\chi$). 
    The latter cannot be resonance annihilation, but its relevant coupling $c_{\phi \phi h}$ and $c_{\phi \phi h h}$ are not suppressed as aforementioned.

    Similar to the SM final state, the annihilation cross-section to a pair of $\phi$ also has a velocity suppression if $c_{p}=0$, 
    while the case $c_{s}=0$ includes both $s$-wave and $p$-wave contributions.
    Note that the $s$-wave cross-section in this channel can be obtained by setting $\left|c_{s}\right| \approx \left|c_{p}\right|$. 
    Qualitatively, the complete annihilation cross-section of this channel exhibits $p$-wave dominant contribution at high velocities, 
    while it becomes $s$-wave dominant if velocity is small. 
\end{itemize}

\subsubsection{Decay width of the scalars}

\begin{figure}[ht!]
	\centering
	\includegraphics[width=12cm,height=9cm]{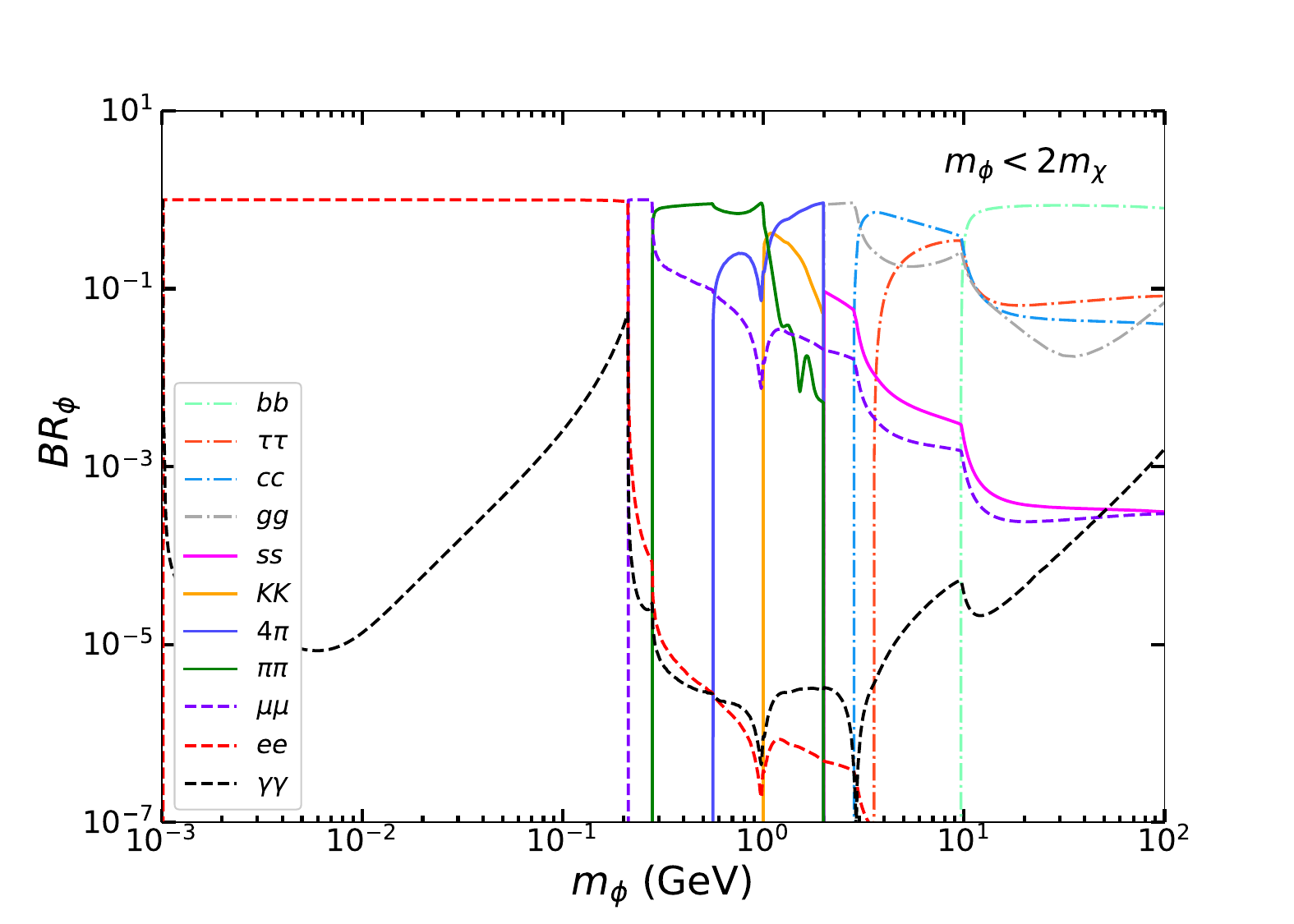}
\caption{The main decay branching ratio of mediator $\phi$. 
Here, we let $m_\chi>m_\phi$, thus $\phi$ does not decay into a DM pair.}
	\label{fig:result_BR}
\end{figure}

Because of $h-\phi$ mixing, the $\phi$ decay width of $\phi\to {\rm SM}+{\rm SM}$ can be directly rewritten as 
\begin{equation}
\Gamma \left ( \phi \to {\rm SM}+{\rm SM} \right ) = 
\sin ^{2} \theta\times \left[\Gamma \left ( h  \to{\rm SM}+{\rm SM}\right )\right]_{ m_{h}\to m_{\phi }  }. 
\label{eq:phidecay}
\end{equation}
If $\phi$ decays to a pair of the SM fermions $f\bar{f}$, the decay width is 
\begin{eqnarray}
	\Gamma \left ( \phi \to f\bar{f} \right ) = \sin ^{2}  \theta \times\frac{m_{f}^{2} m_{\phi } }{8\pi \upsilon _{H}^{2} } \left ( 1-\frac{4m_{f}^{2} }{m_{\phi }^{2} }  \right )^{3/2}.  
\end{eqnarray}
Fig.~\ref{fig:result_BR} depicts the decay branching ratio of several major channels of the mediator $\phi$, where
the mediator mass is less than twice the DM mass~\cite{Binder:2022pmf,Winkler:2018qyg}.
For $m_\phi < 2m_\mu$, the decay channel of $\phi$ will be dominated by electron and photon pairs. 
Instead, more hadron channels can open as the $m_\phi$ increases.
Moreover, for invisible $\phi$ decay ($\phi \to \chi \chi$), the partial decay width can be calculated as   
\begin{eqnarray}
	\Gamma(\phi \to \chi \chi) =\cos^2 \theta \frac{ m_\phi}{16 \pi} \; \left [ c_s^2\left(1 - \frac{4 m_\chi^2}{m_\phi^2} \right)^{3/2}+c_p^2\left(1 - \frac{4 m_\chi^2}{m_\phi^2} \right)^{1/2}\right ].
	\label{eq: phi to chi chi}
\end{eqnarray}

Similar to $\phi$ decay, the expressions of the SM Higgs decay widths can be obtained by replacing $\cos\theta\to \sin\theta$ and $m_\phi\to m_h$.  
However, when $m_h>m_\phi$, the decay width of $h \to \phi \phi$ is given by
\begin{eqnarray}
    \Gamma(h \to \phi \phi ) = \frac{c_{\phi \phi h}^2}{32 \pi m_h} \left( 1 - \frac{4 m_\phi^2}{m_h^2} \right)^{1/2},
\end{eqnarray}
with the couplings $c_{\phi \phi h}$ given in appendix~\ref{app: scalar interactions}
\subsection{The model parameters}

There are only eight free parameters from the scalar potential, 
namely $\mu_H^2$, $\mu_1^3$, $\mu_\Phi^2$, $\mu_3$, $\lambda_H$, $\lambda_\Phi$, $\lambda_{\Phi H}$, $A_{\Phi H}$. 
In addition, only three parameters ($m_\chi$, $c_s$, $c_p$) are in the dark sector. 
In Sec.~\ref{sec:mass}, we have chosen $m_\phi$ and $\sin\theta$ instead of $A_{\Phi H}$ and $\lambda_{\Phi H}$ as the inputs.  
Similarly, the parameter $\lambda_H$ can be determined by the Higgs mass measurement $m_h \simeq 125\gev$. 
Using the minimum condition with given the vacuum expectation values ($v_H \simeq 246\gev$ and $v_\Phi = 0$), 
we can further reduce two parameters from equations $\mu_H^2 + \lambda_H v_H^2/2 = 0$ and $\mu_1^3 + A_{\Phi H} v_H^2/2 = 0$. 
Finally, we have five free parameters to describe the scalar potential ($m_\phi$, $\sin \theta$, $\mu_\phi^2$, $\mu_3$, $\lambda_\Phi$), 
and three parameters ($m_\chi$, $c_s$, $c_p$) to describe the dark sector. 

Including these eight parameters, our prior ranges are
\begin{eqnarray}
	1~\mev \leq &m_\chi& \leq 30\,{\rm GeV}, \nonumber \\
	-1 \leq &c_p& \leq 1, \nonumber \\
        -1 \leq &c_s& \leq 1, \nonumber \\
	1~\mev \leq & m_\phi & \leq 60\,{\rm GeV}, \nonumber \\
	-\pi/6 \leq & \theta& \leq \pi/6, \nonumber \\
	- 1\,{\rm TeV}^2 \leq &\mu_\Phi^2 & \leq 1\,{\rm TeV}^2, \nonumber \\
	-1 \,{\rm TeV} \leq &\mu_3& \leq 1\,{\rm TeV}, \nonumber \\
	-1 \leq &\lambda_\Phi& \leq 1. 
	\label{eq: the parameters}
\end{eqnarray}
As suggested by Ref.~\cite{Matsumoto:2018acr}, except for turning on the pseudo-scalar coupling $c_p$, 
we set the mixing angle $|\theta| \leq \pi/6$ and enhance the DM and mediator mass lower limits to $1\mev$.  
We conservatively assume those dimensionless couplings $|c_s| \leq 1$, $|c_p| \leq 1$ and $|\lambda_\Phi| \leq 1$. 
Additionally, we can restrict those dimensionful parameters ($m_{\phi}$, $|\mu_\Phi|$, $\mu_3$) to be less than $1\tev$. 
Beyond the $1\tev$ scale, an effective theory shall be applied, but it is not in our interest. 
However, we only consider the light DM mass region $m_\chi<30\gev$ in this work.

\section{Constraints}
\label{sec:Constraints}

In this section, we summarize the likelihoods used in our analysis. 
Particularly, we focus on those updated constraints not included in the previous literature.     
Because $s$-wave annihilating DM, especially for $m_\chi<1\gev$, may be hard to satisfy the relic density and 
the constraints from the CMB power spectrum simultaneously, 
we have to take Planck CMB measurement into account. 
Moreover, the current constraints from colliders and DM direct detection  experiments strongly restrict a large $\sin\theta$, so that 
the DM self-interaction can be enhanced to satisfy the relic density. 
Hence, we also use bullet cluster constraints to prevent such a scenario. 

Except for the vacuum stability criterion for the Higgs potential as implemented in Ref.~\cite{Matsumoto:2018acr}, 
we classify other constraints into four groups: 
(i) the cosmological and astrophysical constraints in Sec.~\ref{sec:cos_astro}, 
(ii) the collider constraints in Sec.~\ref{sec:coll}, 
(iii) the constraints from the CMB power spectrum in Sec.~\ref{sec:cmb}, and 
(iv) the bullet cluster constraints for the DM self-interaction in Sec.~\ref{sec:sidm}.

\subsection{Cosmological and astrophysical constraints}
\label{sec:cos_astro}

\begin{table}[h]
\small
\centering
\begin{tabular}{|c|c|c|c|c|}
\hline\hline
      &  Likelihood & Constraints  \\
\hline\hline
\makecell[c]{ Relic abundance} 
      & Gaussian  &  \makecell[c]{$ \Omega_{\chi}^{\rm exp} h^{2} =0.1193\pm 0.0014$~\cite{Planck:2018vyg};\\ 
      $\sigma_{\rm sys}= 10\%\times \Omega_{\chi}^{\rm th} h^{2}$.} \\
\hline
\makecell[c]{ Equilibrium }
      & Conditions
      &\makecell[c]{ 
      either ($\Gamma^{\rm FO}_{\chi{\rm SM}} \geq H_{\rm FO}$), or\\ 
      ($\Gamma^{\rm FO}_{\phi{\rm SM}} \geq H_{\rm FO}$ and 
      $\Gamma^{\rm FO}_{\chi \phi}\geq H_{\rm FO}$)   } \\
\hline
\makecell[c]{DM direct detection}
& Half Gaussian 
&\makecell[c]{
$ 9 \gev < m_{\phi} < 10\tev$ (LZ~\cite{LZ:2022lsv}),\\ 
$ 3.5 \gev <m_{\phi} < 9\gev$ (PANDAX-4T~\cite{PandaX-4T:2021bab}),\\
$ 60\mev <m_{\phi} < 5\gev$ (DarkSide~\cite{DarkSide-50:2023fcw}).
}  
\\
\hline
\makecell[c]{$\triangle N_{\rm eff}$}
&Half Gaussian&$\triangle N_{\rm eff}< 0.17$ for 95\% C.L.~\cite{Planck:2018vyg}\\
\hline
BBN
& Conditions  &\makecell[c]{
if ($m_{\phi}\ge 2m_{\pi}$) then $\tau_{\phi}\le 1$~s~\cite{Berger:2016vxi}, \\
if ($m_{\phi}\le 2m_{\pi}$) then $\tau_{\phi}\le 10^{5}$~s~\cite{Krnjaic:2015mbs}.
 }  \\
\hline\hline

\end{tabular}
\caption{\label{Tab:cosmology}
Cosmological and astrophysical conditions and constraints implemented in our likelihoods. 
The interaction rates at the freeze-out are denoted as $\Gamma^{\rm FO}$, 
while the universe expansion at the freeze-out is $H_{\rm FO}$.  
}
\end{table}

In Table~\ref{Tab:cosmology}, we summarize the cosmological and astrophysical constraints of the DM and the mediator used in this work.  
The Planck measurement $\Omega_{\chi}^{\rm exp} h^{2}$ together with $\sigma_{\rm sys}$ mainly determine the shape of parameter space. 
The Boltzmann solver~\texttt{MicrOMEGAs}~\cite{Belanger:2013ywg} is hired to compute the predicted relic density, 
and $10\%$ of theoretical computation $\Omega_{\chi}^{\rm th} h^{2}$ is then taken as the systematic uncertainties $\sigma_{\rm sys}$ for conservative treatment. 
In most model parameter space, the relic density is lower than the measurement, corresponding to the low DM annihilation rate.    
For $m_{\phi} >  m_{\chi}$, the DM annihilation to SM particles is mainly via $s$-channel by the exchange of $\phi$ and $h$.  
However, these processes are suppressed due to small Yukawa couplings and the mixing angle. 
Therefore, only at the resonant region ($2 m_{\chi} \approx m_{\phi}$) the DM annihilation cross-section can be significantly enhanced 
to fulfill the Planck measurement. 
If $m_{\phi} <  m_{\chi}$, DM can annihilate to a pair of mediators, and the allowed parameter space can be more extended. 
There are two possible solutions: $m_\chi\sim m_\phi$ with a large $\sin\theta$ for a kinematic suppression or 
$m_\phi\ll m_\chi$ with a tiny $\sin\theta$.   
We note those DM annihilation mechanisms may result in the temperature of the dark sector being different from the SM sector before the DM freeze-out. 
This implies that the assumption of the thermal DM can be incorrect for some parameter space.   
To maintain the thermal DM assumption, we force the allowed parameter space to obey the conditions given in Table~\ref{Tab:cosmology}. 
If the interaction rate between the dark sector and the SM sector is stronger than the Universe expansion rate at the freeze-out, namely 
$\Gamma^{\rm FO}_{\chi{\rm SM}} \geq H_{\rm FO}$, the temperatures of the two sectors can be still the same. 
Once $\Gamma^{\rm FO}_{\chi{\rm SM}} < H_{\rm FO}$, the thermal equilibrium can be maintained as long as 
the conditions $\Gamma^{\rm FO}_{\phi{\rm SM}} \geq H_{\rm FO}$ and $\Gamma^{\rm FO}_{\chi \phi}\geq H_{\rm FO}$ hold.

The null signal measurement of DM direct detection (DD) experiments provides stringent upper limits on the DM-proton interactions, especially for $m_\chi$ around GeV. 
Due to the different materials of detectors, different experiments are sensitive to different DM mass ranges, 
such as $ 60\mev <m_{\phi} < 5\gev$ for DarkSide-50~\cite{DarkSide-50:2023fcw}, 
$ 3.5 \gev <m_{\phi} < 9\gev$ for PANDAX-4T~\cite{PandaX-4T:2021bab}, and 
$ 9 \gev < m_{\phi} < 10\tev$ for LZ~\cite{LZ:2022lsv}. 
We incorporate these latest upper limits in our likelihood analysis.\footnote{We incorporate experimental cross-section upper limits from various collaborations into our likelihood analysis. The local DM density is typically assumed to be $0.3\gev$cm$^{-3}$ by experimental collaborations. However, a higher value of $0.4\gev$cm$^{-3}$ is frequently used, which leads to stronger exclusion limits in our study.}
Our theoretical predictions for DM-nucleon elastic scattering cross-section $\sigsip$ at tree-level are mainly computed by using \texttt{MicrOMEGAs}. 
In principle, at the tree-level calculation, only the scalar interaction $c_s$ contributes to the cross-section, 
but the pseudo-scalar interaction $c_p$ is velocity suppressed. 
However, in this work, the loop-level cross-section has to be considered if $c_s\approx 0$. 
The expression of the loop-level contribution can be found in Ref.~\cite{Abe:2018emu}, 
\begin{equation}
             \sigma _{\rm SI}^{\rm loop} = c_{p} ^{4} \cos^{2}  \theta \frac{m_{\chi }^{4}m_{N }^{4} f_{N}^{2}c_{\phi\phi h}^{2}    }{16\pi ^{3}v_{H}^{2} m_{h }^{4} (m_{N }+m_{\chi })^{2}    }
             \left[\frac{\partial B_{0}(m_{\chi },m_{\phi },m_{\chi }) }{\partial p^{2} } \right]^{2},
\label{y}
\end{equation}
where $m_{N}$ is the mass of a nucleon and $f_{N}=f_{Tu}+f_{Td}+f_{Ts}+\left(\frac{2}{9}\right)f_{TG}$, 
with $f_{Tu}\simeq 0.0153$, $f_{Td}\simeq0.0191$, $f_{Ts}\simeq0.0447$, and $f_{TG}\simeq0.921$, respectively.
The loop function can be written as 
\begin{equation}
\frac{\partial B_{0} \left ( m_{\chi } ,m_{\phi },m_{\chi }   \right ) }{\partial p^{2} } = 
\int_{0}^{1}\mathrm{d}x\frac{x\left ( 1-x \right ) }{m_{\phi}^{2} x+m_{\chi }^{2}(1-x)^2}, 
\end{equation}
and the coupling $c_{\phi\phi h}$ is given in appendix~\ref{app: scalar interactions}. 
Hence, we consider the tree-level cross-section for $c_s$ coupling while the loop-level cross-section is computed for $c_p$ coupling.

Finally, a light $\phi$ may contribute the relativistic degree of freedom $\Delta N_{\rm eff}$. 
Following Ref.~\cite{Dolgov:2002wy, Planck:2018vyg}, we apply a 95\% upper limit $ N_{\rm eff}<0.17$~\cite{Planck:2013pxb} as a half Gaussian likelihood. 
This upper limit and the relic density measurements can jointly exclude the region $m_\phi$ and $m_\chi$ below a few MeV.  
Moreover, $\phi$ decaying into the SM particles can also spoil the successful BBN history.  
This can put a stringent limit on the decay time $\tau_\phi$. 
In this work, we adopt a conservative way to implement BBN constraints. 
For the light mass $m_{\phi} < 2m_{\pi}$, because $\phi$ can only decay to leptons, we set a condition $\tau_\phi<10^{5}$~s~\cite{Krnjaic:2015mbs}. 
For the heavier mass $m_{\phi} > 2m_{\pi}$, the $\phi$ hadronic decay can alter the BBN history if its lifetime is shorter than one second~\cite{Berger:2016vxi}.

\subsection{Constraints from collider experiments}
\label{sec:coll}

\begin{table}[]
\small
\centering
\begin{tabular}{|c|c|c|c|c|c|}
\hline\hline
      &  $\phi$ signature & Constraints  \\
\hline\hline
\multirow{3}*{ Higgs decay}& Prompt*  &\makecell[c]{
See the upper limits of BR$\left (h \to \phi \phi \right)$BR$\left ( \phi \to ll  \right )^{2}$\\ 
from Fig.~12 of Ref.~\cite{ATLAS:2021ldb} and Fig.~7 of Ref.~\cite{CMS:2017dmg}.}\\  
\cline{2-3}  ~  & Displaced*  
& See Ref.~\cite{ATLAS:2019tkk,Clarke:2015ala} \\\cline{2-3}  ~ 
& Long-lived* &BR$\left (  h \to {\rm inv.} \right )_{\rm BSM}\le 0.145$~\cite{ATLAS:2022yvh}\\
\hline
\multirow{3}*{$B$ decay} 
&Prompt & BR$\left (  B^{\pm} \to K ^{\pm}  \mu^{-} \mu^{+} \right ) \lesssim 3\times 10^{-7}$~\cite{LHCb:2012juf} \\ \cline{2-3}  ~  
&Displaced &\makecell[c]{
(1) $\sin ^{2} \theta \gtrsim2\times 10^{-8}$ for the region 
\\$0.5<m_\phi/\gev<1.5$ and $1<c\tau_\phi/{\rm cm}<20$~\cite{BaBar:2015jvu}\\  
(2) See Fig.~5 of Ref.~\cite{LHCb:2016awg} for details. }\\  \cline{2-3}  ~   
& Long-lived* &\makecell[c]{ $P_{p} $ BR$\left (B^{\pm} \to K ^{\pm} \nu \bar{\nu} \right )=\left ( 2.3\pm 0.7 \right ) \times 10^{-5}$~\cite{Belle-II:2023esi}}\\
\hline
\multirow{3}*{Kaon decay}& 
Prompt &\makecell[c]{
(1) BR$\left (  K^{+} \to \pi ^{+} \mu^{-} \mu
^{+} \right )\le 4\times 10^{-8}$~\cite{NA482:2010zrc}\\ 
(2) BR$\left (  K_{L} \to \pi ^{0} e^{-}e^{+}  \right )\le 2.8\times 10^{-10}$~\cite{KTEV:2000ngj}\\
(3) BR$\left (  K_{L} \to \pi ^{0} \mu^{-} \mu^{+}  \right )\le 3\times 10^{-10}$~\cite{KTeV:2003sls}} \\
\cline{2-3}  ~  & Displaced & CHARM detected events $\gtrsim 2.3$~\cite{Bezrukov:2009yw}\\ 
\cline{2-3}  ~ & Long-lived* & \makecell[c]{
(1) BR$ \left (K_{L} \to \pi^{0} \nu \bar{\nu} \right  )\le 3.0\times 10^{-9} $\cite{KOTO:2018dsc}\\
(2) See BR$\left (  K^{+} \to \pi ^{+} \nu \bar{\nu}\right )$ limits from \\
Fig.~18 of Ref.~\cite{BNL-E949:2009dza} and Fig.~4 of Ref.~\cite{NA62:2020xlg} for details.}\\
\hline\hline
\end{tabular}
\caption{The constraints from collider experiments. 
Most of them are imposed in our likelihood in the same manner as the former work~\cite{Matsumoto:2018acr}, but those starred columns are the updated constraints.}
\label{tab:colldecay}
\end{table}

The model has two new particles $\chi$ and $\phi$. 
Both of them only interact with the SM sector via the Higgs portal. 
The condition $m_\phi<m_h$ allows us to explore the properties of $\phi$ from the precision measurement of Higgs decay at the colliders. 
It also depends on the lifetime of $\phi$, and there are three possible signatures in the detectors.  
The first one is a prompt signal for the case that $\phi$ immediately decays to some SM charge particles after production, 
and these new charged particles can be recognized in the detectors.    
The second potential signal associated with $\phi$ decay is the observation of displaced vertices. 
When $\phi$ propagates a long distance but decays within the range of the detector,
the signal can be treated as a displaced vertex.    
The third method to search for $\phi$ relies on detecting missing energy signals, 
similar to searching for DM production in the final state. 
This approach assumes that $\phi$ has a long enough lifetime to escape from the detectors.

Moreover, in a sizeable $\sin\theta$, $\phi$ can be produced by the meson decay. 
Hence, the precise measurements of Kaon and $B$ meson decay can largely probe the parameter space of $m_\phi$ and $\sin\theta$. 
Like Higgs decay, we divide these meson decay searches into prompt signals, displaced vertices, and missing energy signals.  
Table~\ref{tab:colldecay} lists all the constraints used in this work. 
The majority are the same as the former work~\cite{Matsumoto:2018acr}, but  
the starred columns are the updated likelihoods described in the following paragraphs.

To compute the likelihood of the colliders, we first take the Higgs decay as an example, 
while the likelihoods for $K$ and $B$ decay can be obtained similarly. 
As shown in Table~\ref{tab:colldecay}, we update all the Higgs relevant likelihoods.
\begin{itemize}
    \item Prompt $\phi$ decay: \\
    By assuming instantaneous decay of $\phi$, 
    the ATLAS collaboration has updated to a new upper limit on 
    BR$\left (h \to \phi \phi \right)$BR$\left ( \phi \to ll  \right )^{2}$~\cite{ATLAS:2021ldb}. 
    After taking the surviving probability of $\phi$, namely $\mathcal{P}_{\phi}=\exp[-\sigma/(\gamma\beta c\tau_{\phi  })]$, into account,  
    the limit shall be modified by including $\mathcal{P}_{\phi}$ as 
    \begin{equation}
        \left(1-\mathcal{P}_{\phi}\right)^{2}{\rm Br}\left (h \to \phi \phi \right  )
        {\rm Br}\left ( \phi \to ll  \right )^{2},
    \end{equation}
    with $\sigma =1~\rm mm$ and $\gamma\beta=m_{h}/(2m_{\phi})\sqrt{1-\left (2   m_{\phi} /m_{h}\right )^{2}  } $~\cite{ATLAS:2021ldb}. 

    \item Displaced $\phi$ decay:\\  
The ATLAS collaboration~\cite{ATLAS:2019tkk} have updated 
the constraint on the dark photon decaying into collimated leptons or light hadrons of 
the process $ h\to \gamma _{\rm d}\gamma _{\rm d}+ X$, 
where the $\gamma _{\rm d}$ and $X$ refer to the dark photon and the lightest dark particle. 
With the null result, we can simply modify the limit to $h\to \phi\phi$ process. 
Compared with the 95\% limits of total dark photon events in the detector,  
we reconstruct the allowed limits as  
\begin{equation}
\frac{N^{\phi}_{\rm dec}}{N^{\gamma_{\rm d},95\%}_{\rm dec}}
= \frac{
\mathcal{P}_{\phi}\times {\rm BR}\left ( gg\to h \right )
\times {\rm BR}\left ( h\to \phi\phi \right )
}{
\mathcal{P}^{95\%}_{\gamma_{\rm d}}\times {\rm BR} \left ( gg\to h \right )
\times {\rm BR}\left ( h\to \gamma _{d} \gamma _{d}+ X \right )
}\le 1,
\end{equation}
where $\mathcal{P}^{95\%}_{\gamma_{\rm d}}$ is the 95\% limit of 
the dark photon decaying probability in the beam dump,  
\begin{equation}
 \mathcal{P}^{95\%}_{\gamma_{\rm d}}=
 \exp\left[\frac{-l_{\rm min} }{ 
 \beta_{\gamma_{\rm d}} 
 \gamma_{\gamma_{\rm d}}
 c \tau_{\gamma _{d}  }} \right]  
 -\exp\left[\frac{-l_{\rm max} }{ \beta_{\gamma_{\rm d}} 
 \gamma_{\gamma_{\rm d}}
 c \tau_{\gamma _{d}  }} \right]. 
 \end{equation}
 Similarly, we can compute the $\phi$ decaying probability $\mathcal{P}_{\phi}$ 
 with the replacement $\gamma_{\rm d}\to \phi$. 

 The lifetimes for $\gamma_{\rm d}$ and $\phi$ are $\tau_{\gamma_{\rm d} }$ and $\tau_{\phi}$, respectively. 
 The value of $\tau_{\gamma_{\rm d}}$ is obtained from 
 the contour in Figure 7 of Ref.~\cite{ATLAS:2019tkk}. 
 Here, we fix the branch ratio BR$\left ( h\to \gamma _{d} \gamma _{d}+ X \right )$ to be $20\%$, 
 while BR$\left ( h\to \phi\phi \right )$ is function of $m_\phi$ and $\sin\theta$.  
 The detector parameters are $l_{\rm min}=1.5~{\rm mm}$ and $l_{\rm max}=307~{\rm mm}$.  
 The $\beta_{\gamma_{\rm d}}=E/m_{\gamma_{\rm d}}$, 
Also, the constraints from Ref.~\cite{Clarke:2015ala} are included in our likelihood. 
    \item Long-lived $\phi$:\\ 
In the case of a very long-lived $\phi$, the signature is the complete missing energy, 
namely the Higgs invisible decay branching ration BR$\left ( h\to {\rm inv.} \right )$. 
We use the latest result from the ATLAS collaboration~\cite{ATLAS:2022yvh}, 
which has improved the upper limit to BR$\left ( h\to {\rm inv.} \right ) \le 0.145 $  
at the $90\%$ confidence limit.
\end{itemize}

If $m_\phi$ is smaller than $B$ or Kaon, we can also search for $\phi$ produced by $B$ or Kaon decay. 
As given in Table~\ref{tab:colldecay}, we update the likelihoods of the long-lived $\phi$ signature for both $B$ and Kaon decay, 
while their likelihoods from prompt and displaced $\phi$ decay are the same as previous work~\cite{Matsumoto:2018acr}.

\begin{itemize}
     \item Long-lived $\phi$ from $B$ decay:\\
The Belle II collaboration has released their latest measurement for BR$\left (B^{\pm} \to K ^{\pm} \nu \bar{\nu} \right ) $, 
with a measurement BR$\left (B^{\pm} \to K ^{\pm} \nu \bar{\nu} \right ) = \left ( 2.3\pm 0.7 \right ) \times 10^{-5} $\cite{Belle-II:2023esi}. This constraint is significantly stronger than the previous upper limits provided by Belle and Babar. We interpret it as 
$P_{l}$$  \text{BR}\left (B^{\pm} \to K ^{\pm} \phi\right )$+$P_{p} $ $\text{BR}\left (B^{\pm} \to K ^{\pm} \phi\right )\text{BR}\left ( \phi \to \chi \chi \right ) =  \left ( 2.3\pm 0.7 \right ) \times 10^{-5}$.  Here $P_{l} $ is the probability that the $\phi$ decays outside the detector and the expression can be written as 
\begin{equation}
P_{l} \equiv \frac{1}{2}\int_{0}^{\pi }\mathrm{d}\theta _{\phi  } \sin \theta _{\phi } \exp\left [ - \frac{l_{xy} }{\sin \theta _{\phi } }\frac{1}{\gamma \beta c\tau _{\phi } }   \right ],
\label{2}
\end{equation}
with the size of the detector $l_{xy}\simeq 25~\rm cm$ and the boost factor $\gamma\beta=m_{B}/(2m_{\phi})\sqrt{1-\left(2m_{\phi}/m_{B}\right)^{2}}$.
     \item Long-lived $\phi$ from Kaon decay:\\
The NA62 collaboration has reported a stringent limit on the branching fraction BR$\left (  K^{+} \to \pi ^{+}\phi\right )$~\cite{NA62:2020xlg}. 
We modify this limit as 
$ P_{l}^{'} \text{Br}\left (  K^{+} \to \pi ^{+} \phi \right ) +\left ( 1-P_{l}^{'}  \right ) \text{Br}\left (  K^{+} \to \pi ^{+} \phi \right )\text{Br}\left ( \phi \to \chi \chi  \right ) $\cite{NA62:2020xlg}, where the probability of mediator decays outside the detector is $P_{l}^{'}=e^{-l_{z} /c\gamma\beta \tau_{\phi}}$  
with the size of detector $l_{z}=65~\rm cm$ and the boost factor $\gamma\beta =37.5~\gev/m_{\phi}$. 
The upper limits on the branch fraction at 90\% can be found in Fig.~9 of Ref.~\cite{NA62:2020xlg}. 
However, the constraint on the same branching fraction provided by the E949 collaboration\cite{BNL-E949:2009dza}, 
as mentioned in Ref.~\cite{Matsumoto:2018acr}, is more stringent compared to the updated constraint from the NA62 collaboration within certain mass intervals. Therefore, we include this constraint in our analysis as well.

On the other hand, the neutral Kaon decay, $K_{L } \to \pi ^{0 } + $missing, is also used to find such a very long-lived mediator. However, the SM process, $K_{L } \to \pi ^{0 } + \nu \bar{\nu}$, is a background against the signal. At present, the KOTO experiment put a stringent constraint on the branching fraction as $  P_{l} \text{Br}\left (  K_{L} \to \pi ^{0} \phi \right ) +\left ( 1-P_{l}  \right ) \text{Br}\left (  K_{L} \to \pi ^{0} \phi \right )\text{Br}\left ( \phi \to \chi \chi  \right )\le 3.0\times 10^{-9} $. Here the $  P_{l}$ is the same as in Eq.~\ref{2}, with the size of the detector $l_{xy}\simeq 145 \rm cm$ and the boost factor $\gamma \beta\simeq 1$. 

\end{itemize}

All of the above constraints from the charged and neutral Kaon decays are included in our likelihood analysis.

\subsection{CMB constraints}
\label{sec:cmb}

In the early universe, around $600 \lesssim z \lesssim 1100$, DM annihilation into SM particles can inject energy into the gas, affecting its temperature and ionization. This energy injection can modify the CMB matter power spectra, as determined by the precise measurements from the Planck experiment~\cite{Planck:2018vyg}.
That implies that, for thermal sub-GeV DM where the number density is greater than the case of $m_\chi>1\gev$, the $95\%$ upper limit on the $s$-wave annihilation cross-section is $\langle \sigma v \rangle \lesssim 10^{-27}$ cm$^3$ s$^{-1}$.

In our minimal DM model, we consider two annihilation channels.
One involves SM pair production through an s-channel exchange mediator, denoted as $\chi\chi \to \text{SM SM}$, 
while the other channel involves DM annihilating into a pair of light mediators, denoted as $\chi\chi \to \phi\phi$, resulting in four SM particles. 
As the CMB constraints are only sensitive to the energy injection, we adopt the same upper limit at the same $m_\chi$ for both two and four SM final state particles.
The decay width of mediator $\phi$ is given by Eq.~\eqref{eq:phidecay}, and the estimation of $\chi^2$ for CMB is described as a Half-Gaussian distribution
\begin{equation}
\chi_{\text{CMB}}^2=\left[\frac{\sv_{\text{CMB}}}{\sv^{\text{Planck},90\%}_{\text{CMB}}/1.64}\right]^2 ,
\label{eq:chi2_CMB}
\end{equation}
where $\sv^{\text{Planck},90\%}_{\text{CMB}}$ is the upper limits of the cross-section for a given DM mass at $95\%$ confidence level from Planck 2018 data~\cite{Planck:2018vyg}. 
Therefore, we take the DM annihilation cross-section during the CMB period as
\begin{equation}
\sv_{\text{CMB}}=
 \left\{
        \begin{array}{ll}
            \sv_{\text{SMs}}
            & \quad  {\rm for~} \chi \chi\to {\rm SM}\,{\rm SM}, \\
            \sv_{\phi\phi}\times \left[ {\rm BR}(\phi\to {\rm SM}\,{\rm SM})\right]^2
            & \quad  {\rm for~} \chi \chi\to \phi\phi, 
        \end{array}
    \right.
\label{eq:CMBlimits}
\end{equation}
where $\langle \sigma v \rangle$ is the velocity-averaged annihilation cross-section 
\begin{equation}
\left \langle \sigma v  \right \rangle=\int \sigma v f(v)\mathrm{d}v,
~{\rm with~}
f(v)= \frac{v^{2} }{2\sqrt{\pi }v_{\rm CMB}^{3}} e^{-\frac{v^{2} }{4v_{\rm CMB}^{2} } }.      
\label{eq:averaged_sv}
\end{equation}
Here, we adopt the relative velocity distribution $f(v)$ described by the Maxwell-Boltzmann distribution~\cite{Ferrer:2013cla}.  
We employ the numerical code \texttt{CalcHEP}~\cite{Belanger:2013ywg} to calculate the annihilation cross-section $\sigma v$.
The DM velocity at recombination epoch $v_{\rm CMB}$ is given by 
$v_{\rm CMB}^2=3 x\left(T_{\gamma}^{\rm CMB}/m_\chi\right)^2$, where we relate the DM temperature $T_{\chi}^{\rm CMB}$ to $v_{\rm CMB}$ as $T_\chi^{\rm CMB}=m_\chi v_{\rm CMB}^2/3$. 
The dimensionless temperature parameter $x=\max\left[x_F, x_{\rm kd}\right]$ depends on the  
freeze-out temperature parameter $x_F$ obtained by \texttt{micrOMEGAs}~\cite{Belanger:2013ywg} and kinematic decoupling temperature parameter $x_{\rm kd}$. 

If kinematic decoupling occurs after freeze-out ($x_{\rm kd}>x_F$), the value of the parameter $x_{\rm kd}$ is only relevant for CMB constraints in the range $x_F< x_{\rm kd}\le m_\chi/T_{\gamma}^{\rm CMB}$. Solving at least two Boltzmann equations tracking DM temperature and density evolution is necessary to determine $x_{\rm kd}$ accurately~\cite{Binder:2017rgn,Binder:2021bmg}. We would like to comment on the impacts of $x_{\rm kd}$ on CMB constraints for $s$-wave, $p$-wave, and $\phi$-resonance annihilation if $x_F< x_{\rm kd}\le m_\chi/T_{\gamma}^{\rm CMB}$.
\begin{itemize}
    \item $s$-wave annihilation:\\
         The annihilation cross-section is independent of $x$, so the CMB constraints are entirely unrelated to the value of $x_{\rm kd}$.
         
    \item $p$-wave annihilation:\\
         The value of $\left \langle \sigma v  \right \rangle$ strongly depends on $x_{\rm kd}$ but is greatly suppressed by the tiny DM velocity, well below the Planck upper limits.
         
    \item $\phi$-resonance annihilation:\\  
          The impacts of $x_{\rm kd}$ on this annihilation are intricate~\cite{Belanger:2024bro}. 
          Only the annihilation peaks within the range $x_F< x_{\rm kd}\le m_\chi/T_{\gamma}^{\rm CMB}$ can be examined in the Planck CMB data. However, such large values of $x_{\rm kd}$ correspond to an extremely narrow range of $1-4 m_\chi^2/m_\phi^2$, spanning approximately from $10^{-16}$ to $10^{-9}$ for $m_\chi\approx 100\mev$. 
          As this extremely fine-tuned range, its indirect detection signal may not be testable or it can be already ruled out by constraints from relic density and CMB data. Hence, we do not consider this parameter space.
\end{itemize}

\subsection{Self Interaction constraints}
\label{sec:sidm}
One important fact of the MeV scale DM is that the vertex between $\phi$ and the SM particles ($\sin\theta$ part) is suppressed by experimental measurements such as collider searches and DM direct detection. 
However, the vertex between $\phi$ and the DM particles ($\cos\theta$ part) have to be large enough to maintain the right size of $\sv$ and 
correct relic abundance in the early Universe. 
Consequently, the self-scattering cross-section, namely $\sigma(\chi\chi\to\chi\chi)$, will be significantly enhanced. 
This resulting self-scattering cross-section may excess the upper limit given by Bullet Cluster limit~\cite{Randall:2008ppe}, 
\begin{equation}
\sigma_{\chi\chi\to\chi\chi}/m_{\chi}<1.0~{\rm cm^2/g}. 
\label{eq:bulletCluster}
\end{equation}
Here, the incident velocity of DM is set to be $4000$~km/s.  
Hence, we assume the likelihood is a half Gaussian distribution with the central value at zero as a conservative test.  
We compute $\sigma(\chi\chi\to\chi\chi)$ by using \texttt{CalcHEP}, but find that the predicted values are much smaller than the limit.

\section{result}
\label{sec:result}

\begin{figure}[ht!]
	\centering
	\includegraphics[width=8.1cm,height=8.1cm]{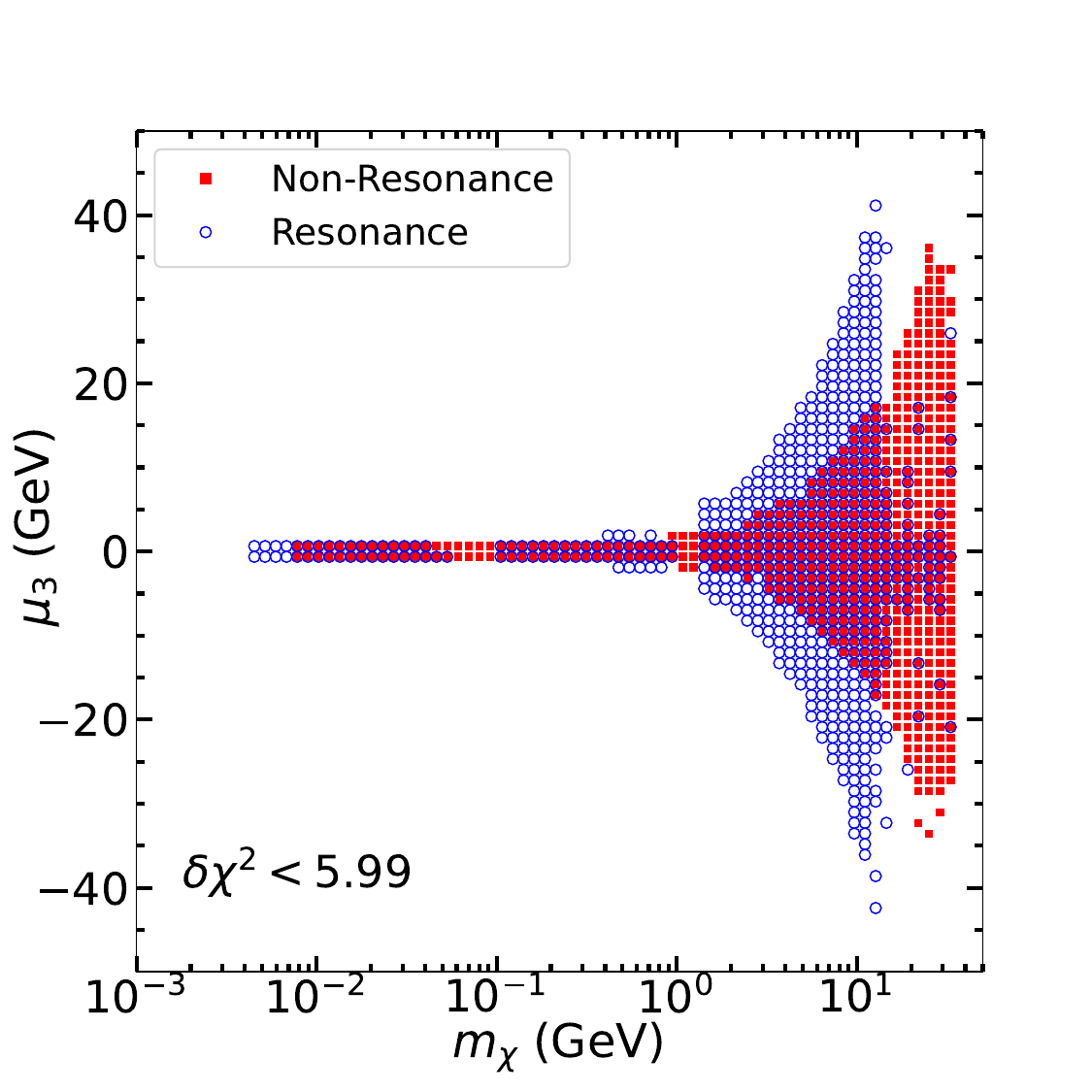}
    \includegraphics[width=8.1cm,height=8.1cm]{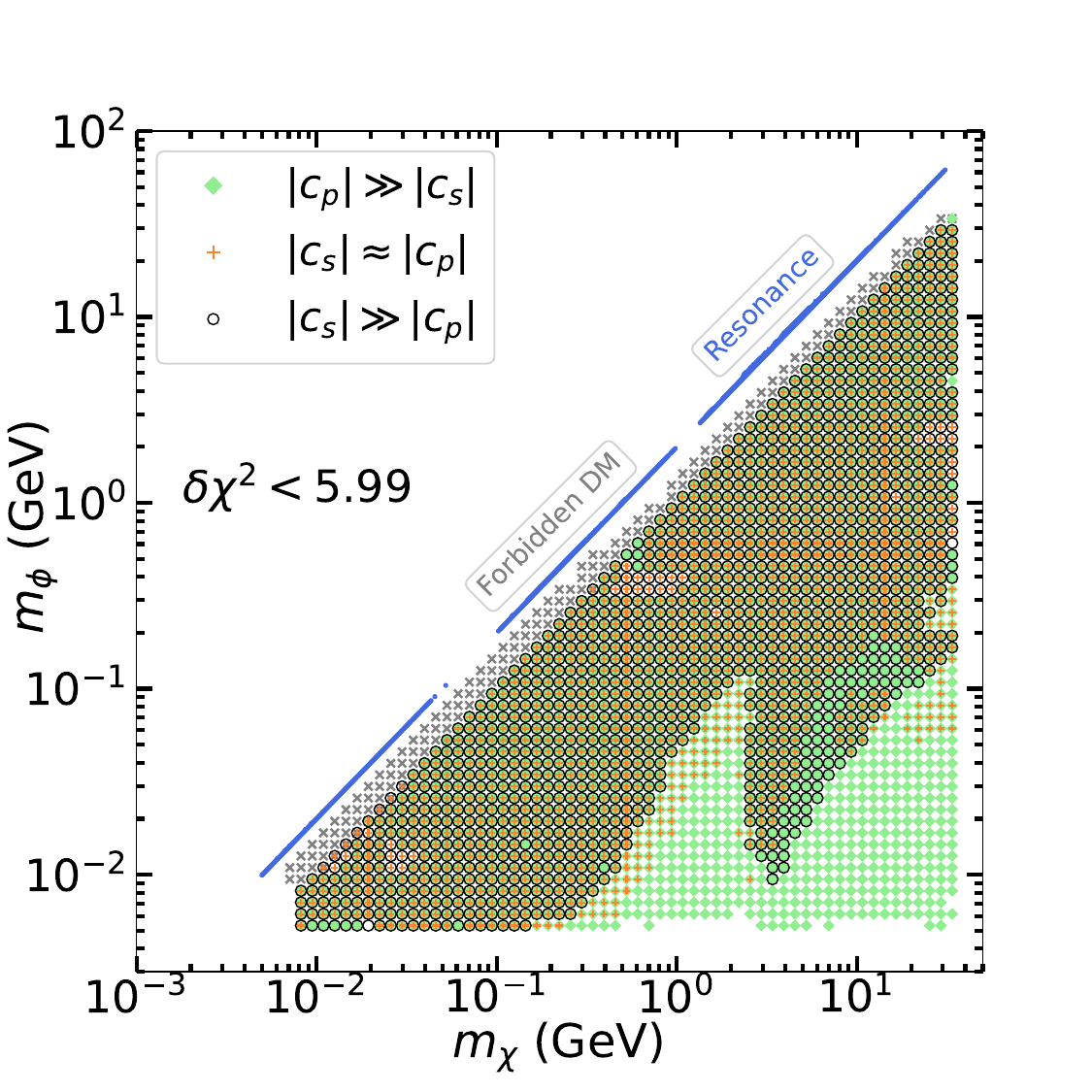} 
\caption{The $95\%$ confidence region in the ($m_\chi$, $\mu_3$) plane (left panel) 
and the ($m_\chi$, $m_\phi$) plane (right panel).  
The colors used for the left panel are the non-resonance annihilations (red solid square) and 
the resonance annihilations (blue unfilled circle).
For the right panel, blue and grey regions represent the resonance and forbidden DM mechanisms, respectively. For the $m_\chi>m_\phi$ cases, the layer for $c_s$ or $c_p$ domination is marked in black circle or green diamond, 
while the scenario of $|c_s|\approx |c_p|$ is indicated in orange plus.
}
	\label{Fig:mx_mu3_mphi}
\end{figure}

In our study, we apply the methodology established in our previous works~\cite{Arhrib:2013ela, Fowlie:2013oua, Tsai:2019eqi, Abdughani:2021oit, Fan:2022dck, Tang:2022pxh}. 
We utilize the likelihood distribution detailed in Sec.~\ref{sec:Constraints} within our Markov Chain Monte Carlo scan. 
By employing \texttt{emcee} package~\cite{Foreman-Mackey:2012any}, 
we conduct 24 Markov chains with several focusing scans to assess good coverage in 
an eight-dimensional parameter space as Eq.~\ref{eq: the parameters}. 
To illustrate our findings, we employ the Profile Likelihood method to eliminate the nuisance parameters and 
project the results onto a two-dimensional plane based on the total three million collected data points.
We assume an approximate Gaussian likelihood for our total likelihood function, 
thus the $95\%$ (2$\sigma$) allowed region is defined by $\delta\chi^2 < 5.99$.

In Fig.~\ref{Fig:mx_mu3_mphi}, we display the $95\%$ confidence region of our total likelihood in two panels, showing the correlations between $m_{\chi}$ and $\mu_{3}$ (left panel) and between $m_{\chi}$ and $m_\phi$ (right panel). The shape of the likelihood distribution is determined by the Planck relic density measurement and thermal equilibrium conditions. In the left panel, we divide the $95\%$ confidence region into two groups: one for non-resonance annihilation (red solid squares) and the other for resonance annihilation (blue unfilled circles). As depicted in the right panel, the relic density constraint favors the non-resonance regions $m_\chi\gtrsim m_\phi$ and the resonance regions $m_\chi \approx m_\phi/2$ (indicated by blue lines). 
For $m_\chi>1\gev$, the resonance region corresponds to a wider range of $\mu_3$ and a heavier $\phi$ compared to the non-resonance region, due to the vacuum stability criterion~\footnote{The parameter correlations can be referenced in Fig.~6 of Ref.~\cite{Matsumoto:2018acr}.}. 
In the mass region $m_\phi<1\gev$ where $m_\chi$ must be less than $\mathcal{O}(\gev)$, 
collider constraints impose an upper limit of $\sin\theta<10^{-2}$, which requires sufficiently small values of $|\mu_3|$ to satisfy the vacuum stability condition.

In the right panel of Fig.~\ref{Fig:mx_mu3_mphi}, 
we divide the $95\%$ confidence region based on the coupling relative size, namely 
$|c_s|>10\times |c_p|$ (black unfilled circles),  $|c_s|<0.1\times |c_p|$ (green solid diamonds), and the rest region $|c_s|\approx |c_p|$ (orange pluses). 
We denote $|c_s|>10\times |c_p|$ as $|c_s| \gg |c_p|$, $0.1\times |c_p|\leq |c_s|\leq 10\times |c_p|$ as $|c_s|\approx |c_p|$, 
and  $|c_s|<0.1\times |c_p|$ as $|c_s| \ll |c_p|$.
We can see that if $|c_s|$ is sufficiently large, in either $|c_s| \gg |c_p|$ or $|c_s| \approx |c_p|$, 
two exclusions appear in $300\mev < m_\chi <30\gev$. 
As discussed in Sec.~\ref{sec:Constraints} regarding the constraints on DM direct detection, 
the spin-independent scattering cross-section of DM-nucleon, dominated by $c_p$, is loop-level suppressed. 
Conversely, if $c_s$ is involved, it contributes at the tree level. 
Therefore, the identified exclusions associated with $c_s$ are mainly imposed by DM direct detection experiments.

We also depict the resonance region $2 m_\chi \approx m_\phi$ with a blue line. 
Two distinct gaps appear around $m_\phi \sim 0.1\gev$ and $m_\phi \sim 2\gev$. 
The exclusion in the former case is a result of the long-lived $\phi$ constraint from Kaon experiments, 
while the latter is ruled out by the Planck relic density and CMB constraints arising from the hadronic $\phi$ decay. 
Moreover, the grey region corresponds to the forbidden DM~\cite{DAgnolo:2015ujb}, namely $m_\phi$ slightly heavier than $m_\chi$. 
In this region, DM particles only annihilate to a pair of $\phi$ before freeze-out, because of the higher DM temperature in the early universe. 
However, the forbidden DM mechanism lacks a visible signal in indirect detection, unless accounting for other specialized DM acceleration mechanisms~\cite{DAgnolo:2015ujb,Delgado:2016umt,DAgnolo:2020mpt,Wojcik:2021xki,Cheng:2022esn}.

\begin{figure}[ht!]
	\centering
    \includegraphics[width=8.1cm,height=8.1cm]{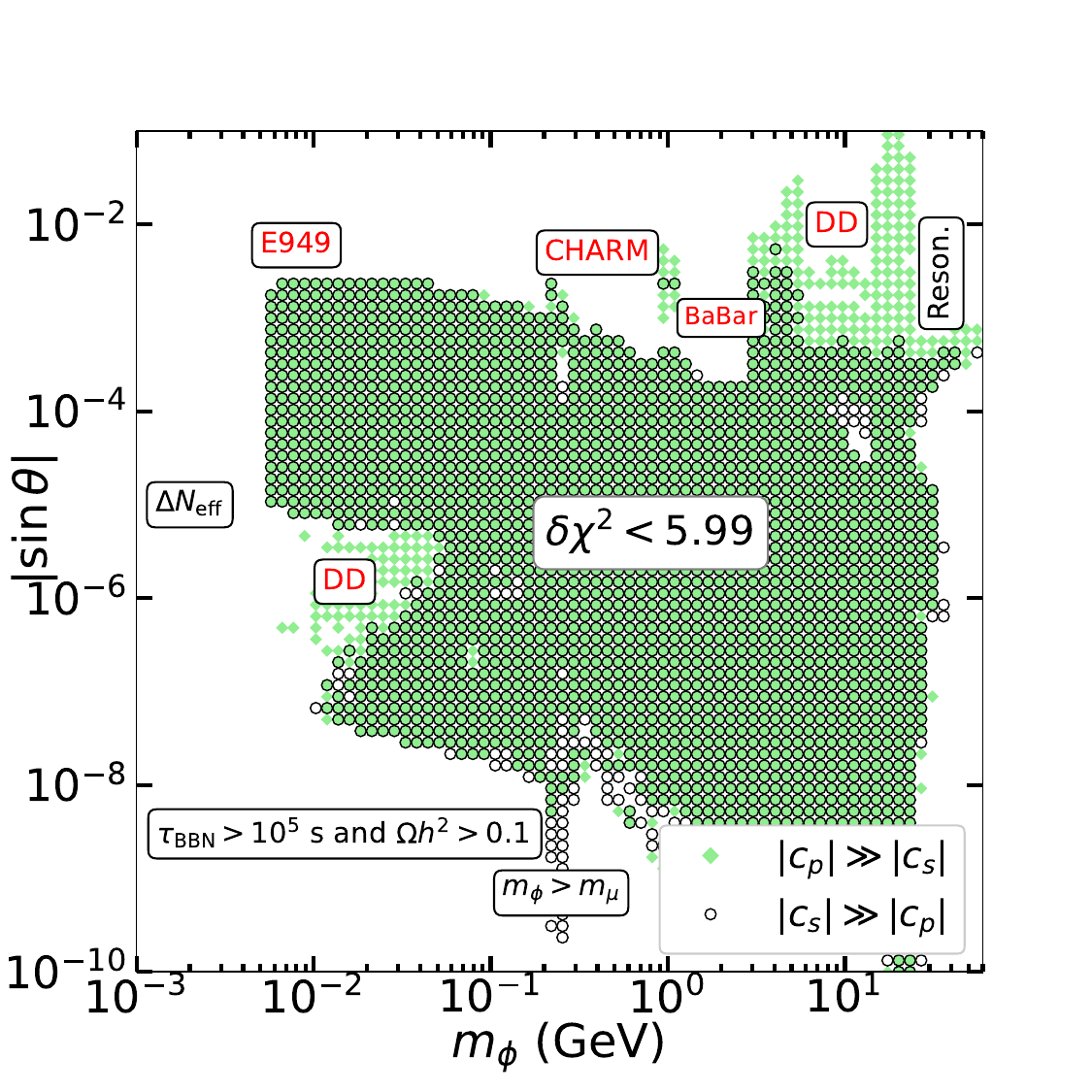}
    \includegraphics[width=8.1cm,height=8.1cm]{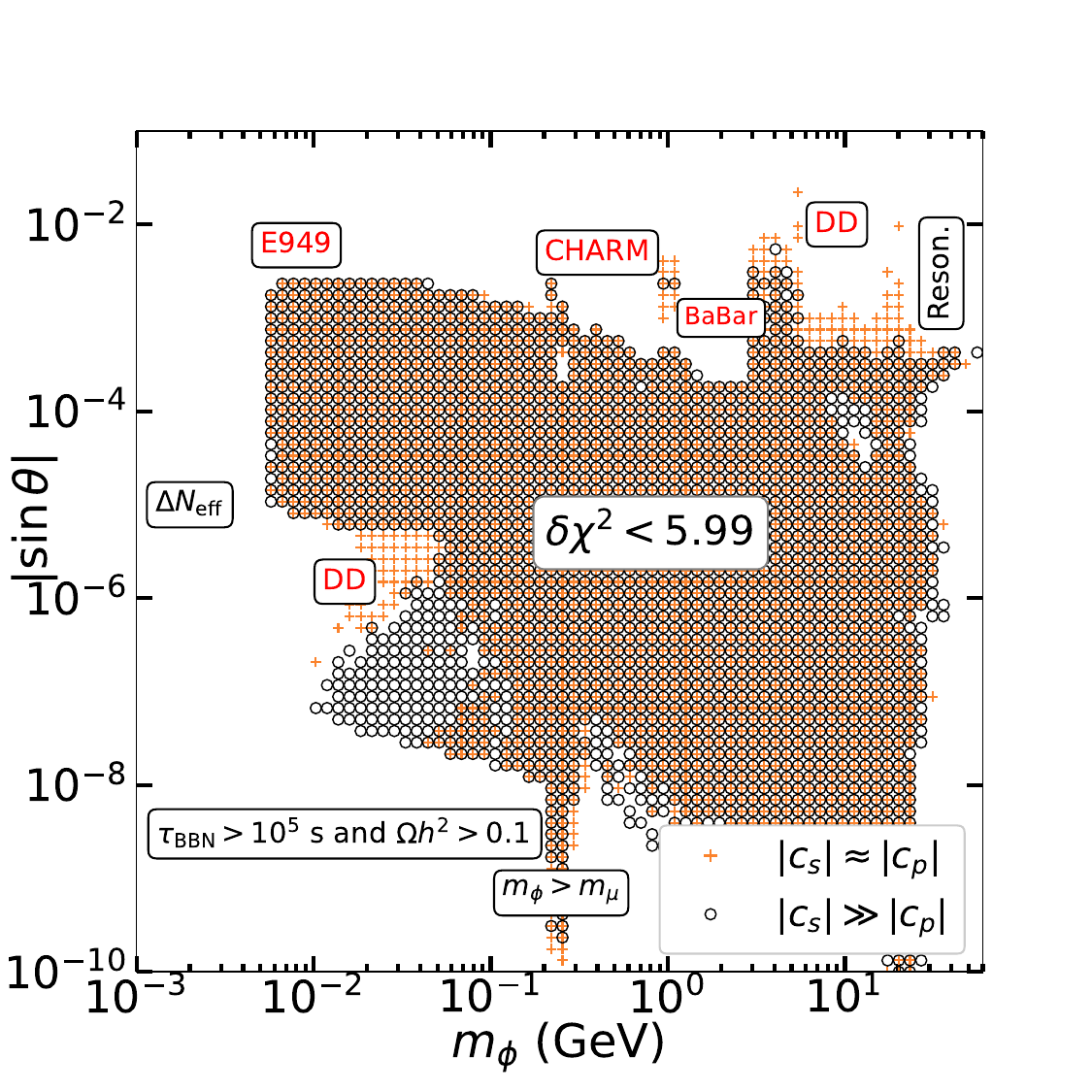}
    \includegraphics[width=8.1cm,height=8.1cm]{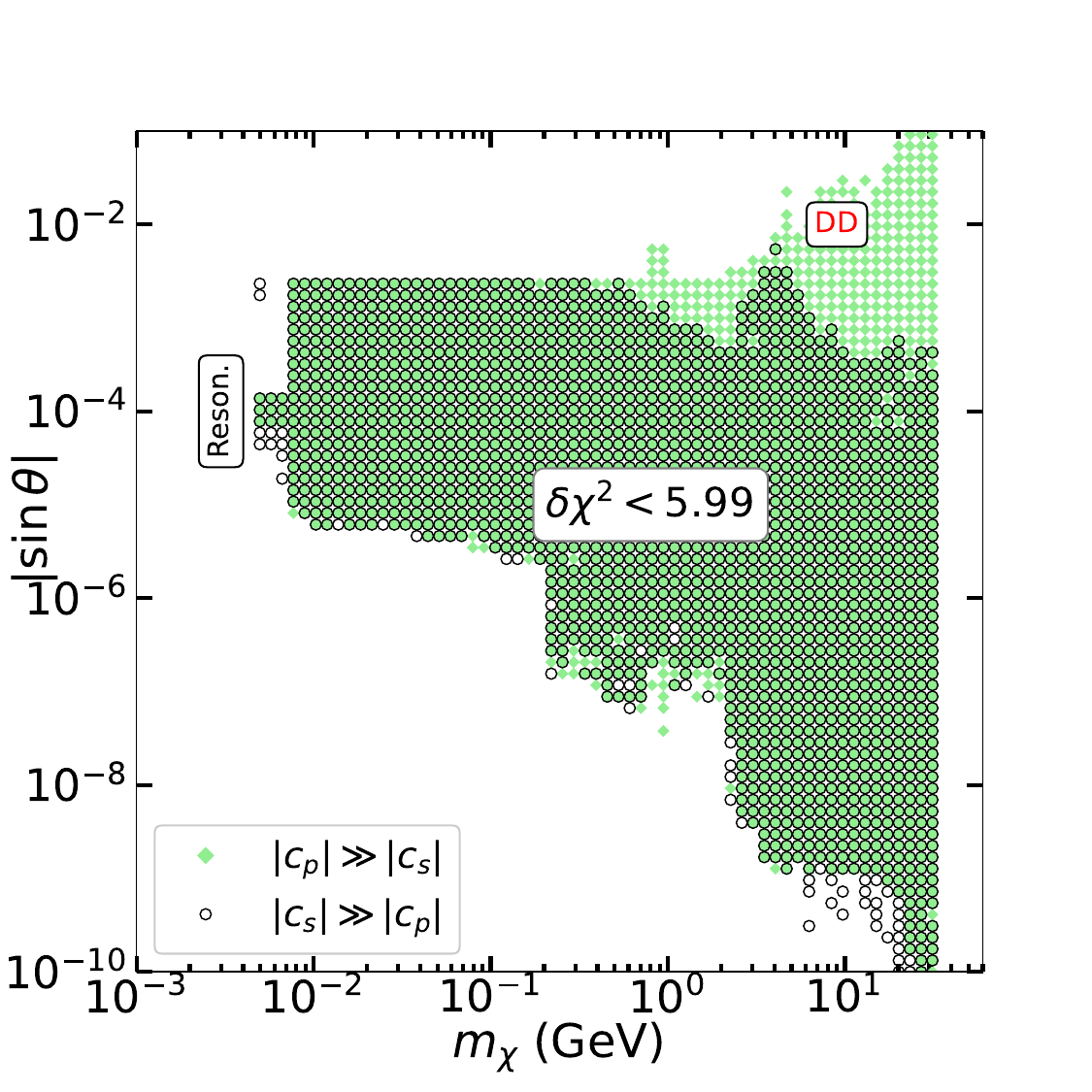}
    \includegraphics[width=8.1cm,height=8.1cm]{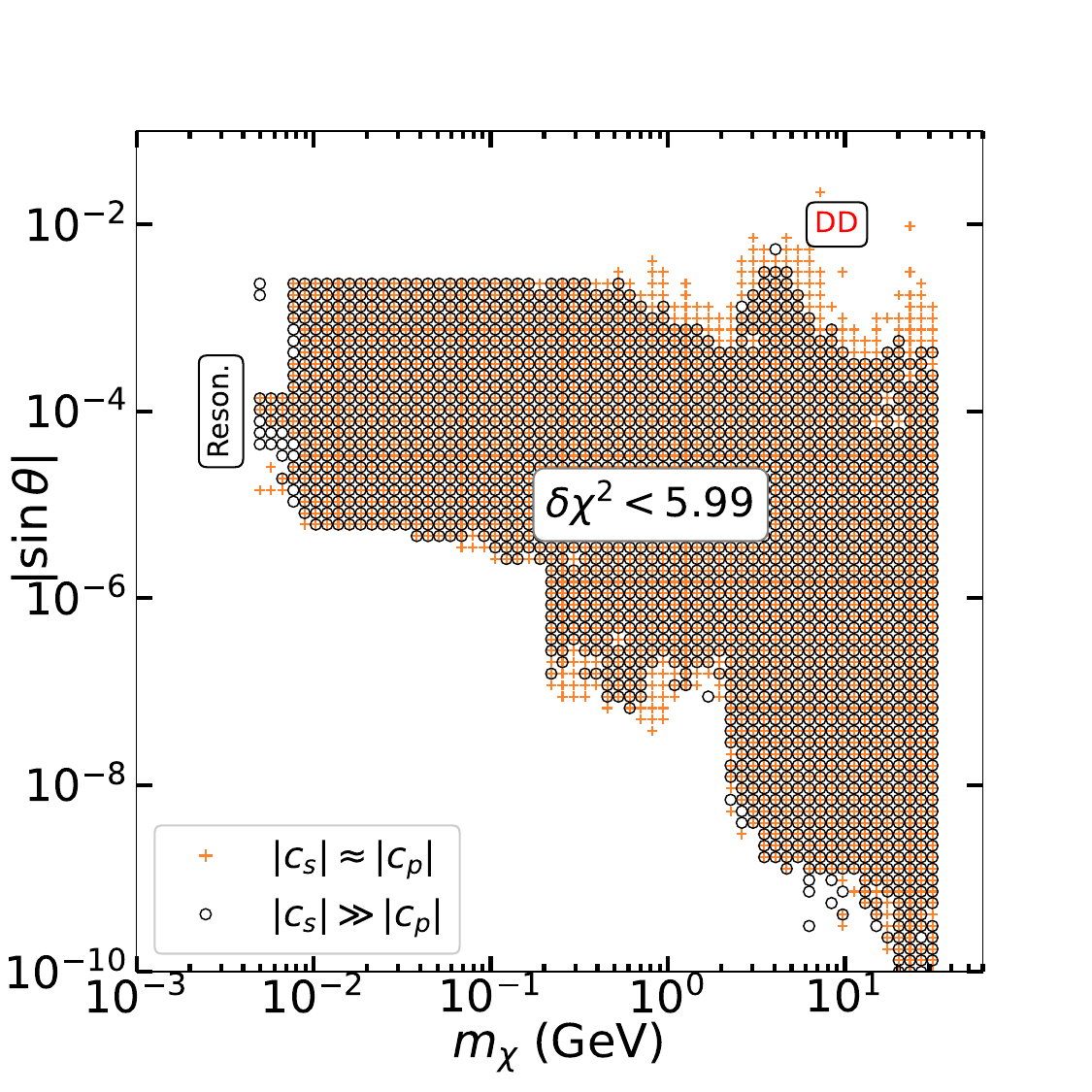}
    \caption{The $95\%$ allowed regions on the panels of $m_\phi$ versus $|\sin{\theta}|$ (two upper panels), 
    and $m_\chi$ versus $|\sin{\theta}|$ (two lower panels).
    The color schemes are the same as the right panel of Fig.~\ref{Fig:mx_mu3_mphi}.
    The red labels indicate the updated constraints from collider and direct detection experiments.}
	\label{fig:sintheta}
\end{figure}

In Fig.~\ref{fig:sintheta}, we present the $2\sigma$ allowed parameter regions on 
the $ \left (  m_{\phi }, \left | \sin \theta  \right |  \right )$ planes (two upper panels) and 
the $ \left (  m_{\chi }, \left | \sin \theta  \right |  \right )$ planes (two lower panels). 
The color scheme is identical to the right panel of Fig.~\ref{Fig:mx_mu3_mphi}. 
Furthermore, we assign appropriate tags of experimental constraints to the exclusion regions. 
Those updated constraints are denoted as "red tags" in comparison to those mentioned in Ref.~\cite{Matsumoto:2018acr}.
We summarize our findings as follows:
\begin{itemize}
\item  For $m_\phi < 500\mev$, Kaon experiments such as CHARM and E949 can probe $\left|\sin\theta\right|\gtrsim 10^{-3}$.
\item  For the range $500\mev \le m_{\phi} \le 5\gev$, the upper limit is determined by $B$ meson experiments. Note that the upper limits for $m_{\phi}$ in the range of $\sim 0.3-4\gev$ exhibit rapid changes. This behavior is attributed to the dominant $\phi$ decay channel, which involves transitions between final states such as $\pi\pi$, $KK$, $4\pi$, $gg$, $c\bar{c}$, and $\tau^{+}\tau^{-}$, as depicted in Fig.~\ref{fig:result_BR}.
\item  In the mass region $m_{\phi} \ge 5\gev$, stringent constraints on $\sin\theta$ are imposed by DM direct detection experiments, particularly in cases where $c_s$ dominates. For cases dominated by $c_p$ (green diamonds), the weaker limit is due to loop suppression. 
Additionally, this region corresponds to $m_\chi > 1\gev$, as illustrated in the two bottom panels. 
\item  For $m_\phi > 30\gev$, only the resonance region $m_\phi \approx 2m_\chi$ survives.
\item  The lower bound on $m_\phi$ is determined by $\Delta N_{\rm eff}$ measurement.

\item  The lower limit on $|\sin\theta|$ is derived from the combined constraints of BBN and the observed DM relic density. The peak around $m_\phi\approx 200\mev$ corresponds to the opening of the $\chi\chi \to \mu^{+}\mu^{-}$ channel. 
In the $|c_p|\gg|c_s|$ scenario, the cross-section of this channel exhibits $s$-wave behavior,  
a result of the simultaneous presence of $\chi \bar{\chi} i \gamma_5 \phi$ and $\Phi |H|^2$ interactions, which break CP symmetry.
Consequently, the parameter space of the $|c_p|\gg|c_s|$ scenario is partially excluded from the constraints imposed by the Planck CMB constraints.

\item The joint contribution of $|c_s| \approx |c_p|$ leads to $s$-wave annihilation of $\chi\chi\to \phi\phi$, in contrast to scenarios where $|c_s| \gg |c_p|$ or $|c_s| \ll |c_p|$, as discussed in Sec.\ref{sec:scalar}.
The $s$-wave annihilation cross-section of this process scales as $\left \langle \sigma v \right \rangle \propto (c_s\times c_p)^2$. 
In the case where $|c_s| \approx |c_p|$ scenario, this results in exclusion for $m_\phi < 100\mev$ and $\left|\sin\theta\right|\lesssim 10^{-6}$, based on the Planck CMB constraints.

\item The parameter region $m_\phi<100\mev$ and $|\sin\theta|\approx 10^{-6}$ with sufficient large contributions from $c_s$ are constrained by DM direct detection.
The requirement for the DM mass to exceed $200\mev$ in this parameter region stems from the need to satisfy kinematic equilibrium conditions.
Despite a suppressed mixing angle, a small $m_\phi$ results in a large scattering cross-section between the DM and nucleons.
On the other hand, the observation of neutrinos from supernova (SN) 1987A presents another potential constraint for this parameter range~\cite{Kamiokande-II:1987idp}.
Refs.~\cite{Dev:2020eam,Balaji:2022noj} show crucial differences in the production rates of scalar and pseudo-scalar interaction from nucleon bremsstrahlung in SN. 
Nevertheless, SNs are complex physical systems, and constraints derived from them are subject to various uncertainties~\cite{Fischer:2016cyd,Mahoney:2017jqk,Tu:2017dhl,Chang:2018rso,Sung:2019xie}.
Therefore, we do not incorporate this constraint into our likelihood analysis.

\end{itemize}

\begin{figure}[htb]
	\centering
    \includegraphics[width=8.1cm,height=8.1cm]{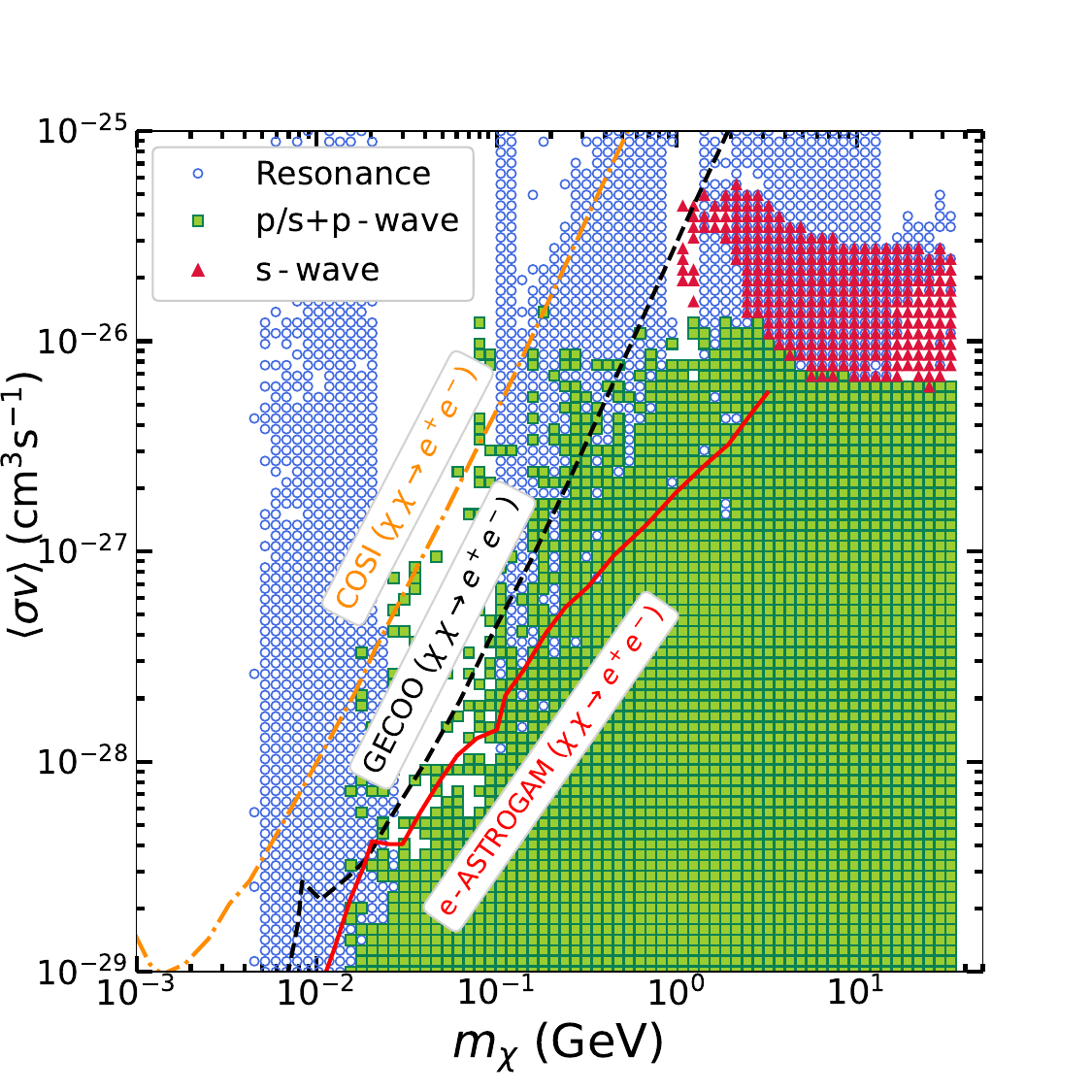}
    \includegraphics[width=8.1cm,height=8.1cm]{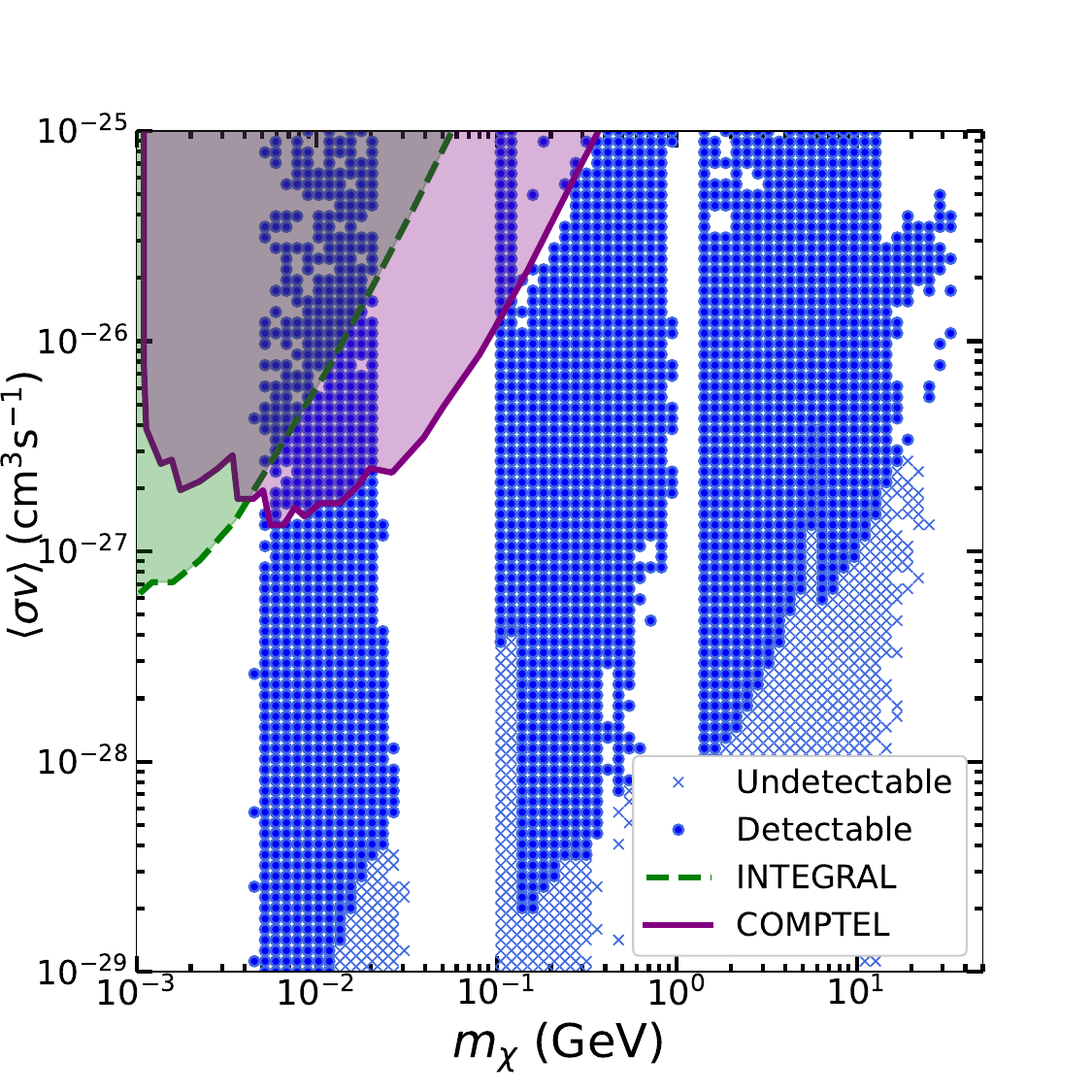}
	\caption{The $95\%$ favored regions of $\langle \sigma v\rangle$ as a function of DM mass $m_\chi$. 
 The red triangles and green squares represent the DM annihilation mechanisms dominated by $s$-wave and $p$ or $s$+$p$-wave, respectively. 
 The $\phi$-resonance annihilation regions are marked with blue circles.
 The solid red line represents the $95\%$ projected upper limit on gamma-ray emission from DM annihilating to $e^+e^-$ in the GC, 
 as determined by e-ASTROGAM~\cite{Strong:2004de}. 
 The DM density profile utilizes the Navarro-Frenk-White (NFW) model. 
 The dashed orange and black lines represent the expectations from COSI~\cite{Caputo:2022dkz} and GECOO~\cite{Coogan:2021rez} for the $\chi\chi\to e^+e^-$ channel, respectively.
 In addition, the green and purple shades in the right panel are from INTEGRAL and COMPTEL limits~\cite{Essig:2013goa}.}  
	\label{fig:mx_sv}
\end{figure}

Finally, we explore the possibilities of testing various DM annihilation mechanisms in upcoming DM indirect detection experiments. 
Fig.\ref{fig:mx_sv} illustrates the $95\%$ favored regions of the annihilation cross-section in the present Universe, 
as a function of $m_\chi$ for all mechanisms (left panel) and specifically for annihilation via $\phi$-resonance (right panel).

In the left panel, we distinguish DM annihilation channels, including $s$-wave (red triangles) and $p$ or $s$+$p$-wave (green squares). 
For $\phi$-resonance annihilation, unfilled blue circles are employed. 
The solid red line represents the $95\%$ projected upper limit of e-ASTROGAM~\cite{Strong:2004de} on gamma-ray emissions from DM annihilating to $e^+e^-$ in the Galactic Center (GC). 
The projected sensitivities for the $\chi\chi\to e^+e^-$ channel from COSI~\cite{Caputo:2022dkz} and GECOO~\cite{Coogan:2021rez} are 
represented by dashed orange and black lines, respectively.

Considering the constraints from the CMB and the relic density, pure DM $s$-wave annihilation only survives at the region $m_\chi>1\gev$. 
Future indirect detection experiments may probe this region.
On the other hand, the DM $p$ and $s$+$p$-wave annihilation cross-section can naturally escape the CMB constraint due to its dependence on velocity squared. 
Therefore, only a small fraction of the parameter space is observable in future indirect detection experiments.

The resonance results present intriguing possibilities for indirect detection. 
In this study, we utilize the Breit-Wigner resonance formula, as outlined in Eq.~(1) of Ref.\cite{Ibe:2008ye}, 
to compute the DM annihilation cross-section for $m_\phi\approx 2 m_\chi$. 
Our resonance result is given in the right panel of Fig.~\ref{fig:mx_sv}. 
Note that the e-ASTROGAM illustrates the detection capability of $\chi\chi \to e^+e^-$, 
but $\phi$ can decay into various channels beyond $e^+e^-$, depending on its mass, as depicted in Fig.~\ref{fig:result_BR}.

To reconstruct the future sensitivity of the e-ASTROGAM for any given annihilation final state, 
we adopt the approach developed in Ref.~\cite{Roszkowski:2012uf} to compute the $95\%$ upper limit on the photon flux induced by DM annihilation. 
Assuming the branching ratio $\text{BR}({\chi\chi\to e^+e^-})$=1, the total flux of DM annihilation-induced $\gamma$-rays 
predicted by a $95\%$ upper limit $\langle \sigma v\rangle_{\text{ann,95}}$ is expressed as
\begin{equation}
\int_{\rm{E_{min}}}^{\rm{E_{max}}} \Phi _{95} \left ( E \right )dE =\frac{N_{e^+e^-}\left \langle   \sigma \upsilon \right \rangle_{e^+e^-,95} J }{8\pi m_{\chi}^{2} }.
\label{eq:photon_flux1}
\end{equation}
Here, \(\text{E}_{\text{min}}\) is the minimum threshold energy of e-ASTROGAM, and \(\text{E}_{\text{max}}\) is the smaller value between the DM mass and the e-ASTROGAM upper threshold energy.
The number $N_{e^+e^-}$ represents the integral of the photon energy distribution per DM annihilation, 
i.e. $N_{e^+e^-}=\int_{\rm{E_{min}}}^{\rm{E_{max}}} \frac{dN_\gamma}{dE_\gamma}(\chi\chi \to e^+e^-)dE_\gamma$. 
The astrophysical factor $J$ can be determined, once the DM halo profile is given.
The bound on the total DM annihilation photons is 
\begin{eqnarray}
	\int_{\rm{E_{min}}}^{\rm{E_{max}}} \Phi _{95} \left ( E \right ) dE=\frac{\left (  \sum_{i} {\rm BR}_{i} N_{i} \right )\left \langle   \sigma \upsilon \right \rangle_{\text{ann},95} J }{8\pi m_{\chi}^{2} }.
  \label{eq:photo_flux2}
\end{eqnarray}
By combining Eq.\ref{eq:photon_flux1} and Eq.\ref{eq:photo_flux2}, we can obtain the total DM annihilation cross-section 
\begin{eqnarray}
  \left \langle   \sigma \upsilon \right \rangle_{\text{ann},95} =
  \frac{\left \langle   \sigma \upsilon \right \rangle_{e ^{+} e^{-}  ,95} N_{e^{+}e^{-}} }{  \sum_{i} {\rm BR}_{i} N_{i} }.
\end{eqnarray}
Here, $\left\langle\sigma \upsilon \right \rangle_{e^+e^-,95}$ is obtained from 
the e-ASTROGAM upper limit of the DM annihilation cross-section (the red solid line in the left panel of Fig.~\ref{fig:mx_sv}).
We employ \texttt{HAZMA}~\cite{Coogan:2019qpu} and \texttt{PPPC4}~\cite{Cirelli:2010xx} to calculate the photon spectrum, 
such as $N_\gamma^{e^+e^-}(E)$, for each DM annihilation channel.
We sum the branching ratios BR$_{i}$ over all the $\phi$ decay channels.

In the right panel of Fig.~\ref{fig:mx_sv}, we depict the expected sensitivity of e-ASTROGAM in detecting DM annihilation through $\phi$-resonance.
The green and purple shaded regions show constraints on the DM annihilation cross-section from the current MeV gamma-ray telescopes, INTEGRAL and COMPTEL~\cite{Essig:2013goa}.
The detectable regions are denoted by blue solid circles, whereas regions marked with blue crosses present challenges in probing with e-ASTROGAM. 
The distinct spike structures at $m_{\chi} \sim 1.05$ GeV and $m_{\chi} \sim 4.18$ GeV correspond to the thresholds of the $\chi \chi \to \mu^{+} \mu^{-}$ and $\chi \chi \to b \bar{b}$ channels, respectively.

\section{Conclusion}
\label{sec:summary}

In this comprehensive study, we investigated a minimal renormalizable DM model, by including a sub-GeV Majorana DM and a singlet scalar particle. 
Building upon the setup of Ref.~\cite{Matsumoto:2018acr}, we extended our analysis by incorporating both scalar and pseudo-scalar interactions ($c_s$ and $c_p$) between the DM and the newly introduced scalar. 
In comparison to the scenario where $c_p$ is absent, the new $c_p$ interaction yields intriguing implications for DM direct and indirect detection.

For DM indirect detection, the interplay between $c_s$ and $c_p$ coupling predicts that DM annihilation cross-section can be dominated by $s$-wave, $p$-wave, or a combination of both ($s$+$p$-wave) contributions.
In the context of DM direct detection, the role of $c_p$ is limited to the loop-correction contribution of DM-nucleon elastic scattering. 
Thanks to nonzero DM-nucleon elastic scattering cross-section, even in the absence of $c_s$, this scenario remains testable.

Our analysis takes into account these new model features and incorporates updated constraints. 
We include constraints from DM direct detection experiments, Planck CMB power spectrum measurements, and collider searches in our likelihood, providing a more comprehensive analysis in light of the latest experimental data.

We have identified a broad parameter space within the $2\sigma$ allowed region, specifically $10\mev < m_\chi \lesssim m_\phi$ and $2 m_\chi\approx m_\phi$
($\phi$-resonance region). 
Additionally, we explore three distinct scenarios: $|c_s|\gg |c_p|$, $|c_s|\ll |c_p|$, and $|c_s|\approx |c_p|$. 
Our findings indicate that the presence of a non-zero pseudo-scalar coupling coefficient effectively alleviates constraints from direct detection. 
However, CMB observations impose rigorous limits on scenarios where $c_p$ dominates, particularly for DM masses lighter than $1\gev$. 
Consequently, the scenario with only $s$-wave annihilation remains viable only for $m_\chi>1\gev$.

Finally, we explored the detectability prospects for this minimal DM model in future indirect detection experiments. 
Our findings suggest challenges for thermal sub-GeV DM annihilating via $s$-wave processes to meet both relic density and CMB constraints, whereas 
$p$-wave annihilation fits these criteria but is hard to be detected by indirect detection experiments. 
Fortunately, a significant portion of the parameter space associated with $\phi$-resonance can be probed by future indirect detection experiments, 
such as e-ASTROGAM.

\section*{Acknowledgments}
We express our gratitude to Mei-Wen Yang for her invaluable numerical support. 
YST is supported by the National Key Research and Development Program of China
(No. 2022YFF0503304), and the Project for Young Scientists in Basic Research of the Chinese Academy of Sciences 
(No. YSBR-092). LW is supported by the National Natural Science Foundation of China (NNSFC)  No. 12275134, No. 12335005, and No. 12147228.
SM is supported by Grant-in-Aid for Scientific Research from the MEXT, Japan (20H01895, 20H00153, 19H05810, 18H05542, JPJSCCA20200002) and by World Premier International Research Center Initiative (WPI), MEXT, Japan (Kavli IPMU). 

\newpage
\appendix

\section{Scalar Interactions}
\label{app: scalar interactions}

Here we show the explicit expression for the scalar interactions 
between $h$ and $\phi$.
\begin{eqnarray}
	c_{h h h} &=& 3 \lambda_H v_H c_\theta^3 - 3 A_{\Phi H} c_\theta^2 s_\theta - \mu_3 s_\theta^3
	+ 3 \lambda_{\Phi H} v_H s_\theta^2 c_\theta,
	\nonumber \\
	c_{\phi h h} &=& 3 \lambda_H v_H c_\theta^2 s_\theta + A_{\Phi H} (c_\theta^3 - 2 c_\theta s_\theta^2 )
	+ \mu_3 c_\theta s_\theta^2 + \lambda_{\Phi H} v_H ( s_\theta^3 - 2c_\theta^2 s_\theta),
	\nonumber \\
	c_{\phi \phi h} &=& 3 \lambda_H v_H c_\theta s_\theta^2 + A_{\Phi H } (2c_\theta^2 s_\theta - s_\theta^3)
	- \mu_3 c_\theta^2 s_\theta + \lambda_{\Phi H} v_H (c^3_\theta - 2 c_\theta s_\theta^2),
	\nonumber \\
	c_{\phi \phi \phi} &=& 3 \lambda_H v_H s^3_\theta + 3 A_{\Phi H} c_\theta s^2_\theta + \mu_3 c_\theta^3
	+ 3 \lambda_{\Phi H} v_H c_\theta^2 s_\theta,
	\nonumber \\
	c_{h h h h} &=& 3 \lambda_H c_\theta^4 + 6 \lambda_{\Phi H} c_\theta^2 s_\theta^2 + \lambda_\Phi s_\theta^4,
	\nonumber \\
	c_{\phi h h h} &=& 3 \lambda_H c_\theta^3 s_\theta - 3 \lambda_{\Phi H} (c_\theta^3 s_\theta - c_\theta s_\theta^3)
	- \lambda_\Phi c_\theta s_\theta^3,
	\nonumber \\
	c_{\phi \phi h h} &=& 3 \lambda_H c_\theta^2 s_\theta^2
	+ \lambda_{\Phi H} (c_\theta^4 - 4 c_\theta^2 s_\theta^2 + s_\theta^4 ) + \lambda_\Phi c_\theta^2 s_\theta^2,
	\nonumber \\
	c_{\phi \phi \phi h} &=& 3 \lambda_H c_\theta s_\theta^3 + 3 \lambda_{\Phi H} ( c_\theta^3 s_\theta - c_\theta s_\theta^3 )
	- \lambda_\Phi c_\theta^3 s_\theta,
	\nonumber \\
	c_{\phi \phi \phi \phi} &=& 3 \lambda_H s_\theta^4 + 6 \lambda_{\Phi H} c_\theta^2 s_\theta^2 + \lambda_\Phi c_\theta^4,
\end{eqnarray}
where parameters $s_\theta$ and $c_\theta$ are defined as $s_\theta \equiv \sin\theta$ and $c_\theta \equiv \cos\theta$, respectively.


\end{document}